\documentclass[fleqn,usenatbib]{mnras}

\usepackage{times}

\usepackage[T1]{fontenc}
\usepackage{hyperref}
\usepackage[utf8]{inputenc}
\usepackage{wrapfig}
\usepackage{booktabs}
\usepackage{longtable}
\usepackage{float}
\usepackage{rotating}
\usepackage{lscape}
\usepackage{lipsum}
\usepackage{pdflscape}
\usepackage{afterpage}
\usepackage{makecell}

\DeclareRobustCommand{\VAN}[3]{#2}
\let\VANthebibliography\thebibliography
\def\thebibliography{\DeclareRobustCommand{\VAN}[3]{##3}\VANthebibliography}

\usepackage{natbib}
\usepackage{graphicx}	
\usepackage{epstopdf}
\usepackage{amsmath}	
\usepackage{amssymb}	
\usepackage{amsfonts}
\usepackage{multirow}

\title[Ionization and electron density gradients in reflection spectra]{Impact of ionization and electron density gradients in X-ray reflection spectroscopy measurements}

\author[Mall et al.]{Gitika~Mall,$^{1}$
Ashutosh~Tripathi,$^{1}$
Askar~B.~Abdikamalov,$^{1,2,3}$
and Cosimo~Bambi$^{1}$\thanks{Email: \href{mailto:bambi@fudan.edu.cn}{bambi@fudan.edu.cn}}
\\
$^{1}$Center for Field Theory and Particle Physics and Department of Physics, Fudan University, 200438 Shanghai, China\\
$^{2}$Ulugh Beg Astronomical Institute, Tashkent 100052, Uzbekistan\\
$^{3}$Institute of Fundamental and Applied Research, National Research University TIIAME, Tashkent 100000, Uzbekistan
}

\date{Accepted XXX. Received YYY; in original form ZZZ}

\pubyear{2022}

\begin{document}
\label{firstpage}
\pagerange{\pageref{firstpage}--\pageref{lastpage}}
\maketitle

\begin{abstract}
The models currently used for the analysis of the reflection spectra of black holes usually assume a disk with constant ionization and electron density. However, there is some debate on the impact of these assumptions on the estimate of the properties of the sources, in particular when the fits suggest very steep emissivity profiles in the inner part of the accretion disk. In this work, we re-analyze a selected set of high-quality \textsl{NuSTAR} and \textsl{Suzaku} data of Galactic black holes and we fit the reflection component with three different models: {\tt relxill\_nk}, in which the ionization parameter and the electron density are constant, {\tt relxillion\_nk}, where the electron density is still constant but the ionization profile is described by a power law, and {\tt relxilldgrad\_nk}, where the electron density profile is described by a power law and the ionization profile is calculated self-consistently from the electron density and the emissivity. While {\tt relxillion\_nk} can fit the data better, we do not find any substantial difference in the estimate of the properties of the sources among the three models. Our conclusion is that models with constant electron density and ionization parameter are probably sufficient, in most cases, to fit the currently available X-ray data of accreting black holes. 
\end{abstract}




\section{Introduction}\label{sec:Intro}

Blurred reflection features are commonly observed in the X-ray spectra of accreting black holes~\citep{1989MNRAS.238..729F,1995Natur.375..659T,1997ApJ...477..602N}. They are thought to be produced by illumination of a cold accretion disk by a hot corona~\citep{1995MNRAS.277L..11F,2003PhR...377..389R,2013Natur.494..449R,2021SSRv..217...65B}. The process can be briefly described as follows. The black hole is accreting from a geometrically thin and optically thick disk. The thermal spectrum of the accretion disk turns out to be peaked in the soft X-ray band (0.1-10~keV) in the case of stellar-mass black holes and in the UV band (1-100~eV) for supermassive black holes. The corona is some hotter plasma ($T_{\rm c} \sim 100$~keV) near the black hole and the inner part of the accretion disk. The corona may be some hot atmosphere above the accretion disk, the hot flow in the plunging region between the inner edge of the disk and the black hole, the base of the jet, etc. Some thermal photons of the accretion disk can inverse Compton scatter off free electrons in the corona. The resulting spectrum can be usually approximated well by a power law with photon index $\Gamma \approx 1$-3 and exponential high-energy cutoff $E_{\rm cut} \approx 2$-3~$T_{\rm c}$. A fraction of Comptonized photons can illuminate the disk and interact with the material of the disk: Compton scattering and absorption followed by fluorescent emission generate the reflection spectrum.

In the rest-frame of the gas in the disk, the reflection spectrum presents narrow fluorescent emission lines in the soft X-ray band and a Compton hump peaked at 20-30~keV~\citep{2005MNRAS.358..211R,2010ApJ...718..695G}. The most prominent emission line is usually the iron K$\alpha$ complex, which is at 6.4~keV in the case of neutral or weakly ionized iron and can shift up to 6.97~keV in the case of H-like iron ions. Far from the source, the fluorescent emission lines appear broadened and skewed as a result of relativistic effects in the strong gravity region around the black hole: gravitational redshift, Doppler boosting, and light bending~\citep{1989MNRAS.238..729F,1991ApJ...376...90L,2010MNRAS.409.1534D,2017bhlt.book.....B}. The analysis of these broadened reflection features can potentially be a powerful tool for studying the properties of the inner part of the accretion disk, measuring black hole spins, and even testing fundamental physics~\citep{2013mams.book.....B,2013ApJ...773...57J,2014SSRv..183..277R,2017RvMP...89b5001B,2021SSRv..217...65B}.

Models for the analysis of the reflection spectra of accreting black holes have been significantly improved in the past decade~\citep{2013MNRAS.430.1694D,2013ApJ...768..146G,2014ApJ...782...76G}; for a review, see~\citet{2021SSRv..217...65B}. However, they still rely on a number of simplifications that may introduce unacceptably large systematic uncertainties in the final measurements of the properties of the sources. It is thus crucial to understand well the systematic uncertainties of the theoretical models, as well as to develop more and more sophisticated theoretical models, in order to obtain precise and accurate measurements~\citep{2008ApJ...675.1048R,2020PhRvD.101d3010Z,2020PhRvD.101l3014C,2020ApJ...895...61R,2021ApJ...910...49R,2020arXiv201207469R,2020PhRvD.102j3009T,2021ApJ...913..129T}. Otherwise, with the possibility of analyzing higher and higher quality spectra, there is the risk to get very precise but not very accurate measurements of accreting black holes, which would nullify the efforts to design and launch more powerful X-ray observatories.

Current analyses of the reflection spectra of accreting black holes normally employ reflection models with accretion disks with constant ionization and electron density. The ionization parameter is defined as
\begin{eqnarray}\label{eq-xi}
\xi = \frac{4 \pi F_X}{n_{\rm e}} \, ,
\end{eqnarray}
where $F_X$ is the X-ray flux from the corona illuminating the disk and $n_{\rm e}$ is the electron density of the disk. The radial profile of the X-ray flux $F_X$ is determined by the coronal geometry~\citep{2012MNRAS.424.1284W,2013MNRAS.430.1694D}. Compact coronae very close to their black hole can naturally produce very steep X-ray flux in the inner part of the accretion disk as a result of light bending in the strong gravity region near the compact object~\citep{1996MNRAS.282L..53M,2013MNRAS.430.1694D,2020arXiv201207469R}. The radial profile of the electron density $n_{\rm e}$ depends on the properties of the accretion disk, but it is normally a function that moderately decreases as the radial coordinate increases.

Both the ionization parameter $\xi$ and the electron density $n_{\rm e}$ affect the reflection spectrum in the rest-frame of the gas and are not expected to be constant over radii, which is instead the typical assumption in the analysis of reflection spectra. The assumption of constant ionization parameter and the electron density is certainly a simplification of the models, but it is also motivated by the fact that the strong light bending near the black hole can focus the Comptonized photons from the corona on a quite small portion of the inner part of the accretion disk, which we may thus be approximated well with a one-ionization region. On the other hand, very steep radial profiles of the X-ray flux $F_X$ should lead to very steep radial profiles of the ionization parameter $\xi$, which has quite a strong impact on the shape of the reflection spectrum. \citet{2012A&A...545A.106S} and \citet{2019MNRAS.485..239K} found that employing reflection models with constant ionization profile may lead to overestimate the steepness of the emissivity profile of the inner part of the accretion disk and get inaccurate black hole spin measurements \citep[see also][]{2020MNRAS.492..405S}.
\citet{2022MNRAS.512..761W} found that an ionization gradient is required to model the broadband (0.3-50~keV) reflection spectrum of the Seyfert galaxy I~Zwicky~1.

The aim of our work is to explore whether observations require reflection models with non-constant ionization parameter $\xi$ and electron density $n_{\rm e}$. To do this, we select three high quality spectra of Galactic black holes and we fit every spectrum with three reflection models: {\tt relxill\_nk}~\citep{Bambi_2017,2019ApJ...878...91A,2020ApJ...899...80A}, in which $\xi$ and $n_{\rm e}$ are constant, {\tt relxillion\_nk}~\citep{2021PhRvD.103j3023A}, in which $n_{\rm e}$ is constant but $\xi$ has a radial profile described by a power law, and {\tt relxilldgrad\_nk}~\citep{2021ApJ...923..175A}, where $n_{\rm e}$ has a radial profile described by a power law and $\xi$ is calculated from Eq.~(\ref{eq-xi}) assuming that $F_X$ is proportional to the emissivity profile of the reflection spectrum. We find that {\tt relxillion\_nk} can usually provide a better fit, but we do not find any substantial difference in the estimate of the parameters of the sources among the three different models. Such a conclusion confirms the result found in \citet{2021ApJ...923..175A}, where we presented {\tt relxilldgrad\_nk} and we analyzed a \textsl{NuSTAR} spectrum of the Galactic black hole in EXO~1846--031 with the three reflection models, finding that {\tt relxillion\_nk} fits the data better, but the three models provide consistent measurements of the parameters of the source.

The manuscript is organized as follows. In Section~\ref{sec:m}, we briefly review the three reflection models used in our work, pointing out their main differences. In Section~\ref{sec:o}, we select the observations for our study. In Section~\ref{sec:dr} and in Section~\ref{sec:sa}, we present, respectively, the data reduction and the spectral analysis of the selected observations. Discussion of the results and conclusions are reported in Section~\ref{sec:dc}.


\section{Reflection models}\label{sec:m}

In our study, we employ three different reflection models: {\tt relxill\_nk}, {\tt relxillion\_nk}, and {\tt relxilldgrad\_nk}. In this section we clarify the specific properties of each model.

{\tt relxill\_nk}~\citep{Bambi_2017,2019ApJ...878...91A,2020ApJ...899...80A} was developed as an extension of the {\tt relxill} package \citep{2013MNRAS.430.1694D,2013ApJ...768..146G,2014ApJ...782...76G} to non-Kerr spacetimes. In the present work, we will use {\tt relxill\_nk} with the deformation parameter frozen to zero to impose that the spacetime is described by the Kerr metric. With such a choice, {\tt relxill\_nk} formally reduces to the {\tt relxill} model, but it reads a different table to include all relativistic effects in the final spectra and has some different subroutines, which are the same as in the other two models used in our work. We thus use {\tt relxill\_nk} instead of {\tt relxill} in order to be sure that any difference in the fits is only due to the assumptions on the ionization and electron density profiles. The ionization parameter $\xi$ is assumed to have the same value over the whole disk and is left free in the fits. The electron density $n_{\rm e}$ is frozen to $10^{15}$~cm$^{-3}$ over the whole disk since {\tt relxill\_nk} uses the {\tt xillver} table~\citep{2013ApJ...768..146G,2014ApJ...782...76G} for the reflection spectrum in the rest-frame of the gas.

\begin{figure*}
	\includegraphics[width=0.95\linewidth,trim=2.5cm 0.0cm 3.0cm 0.0cm,clip]{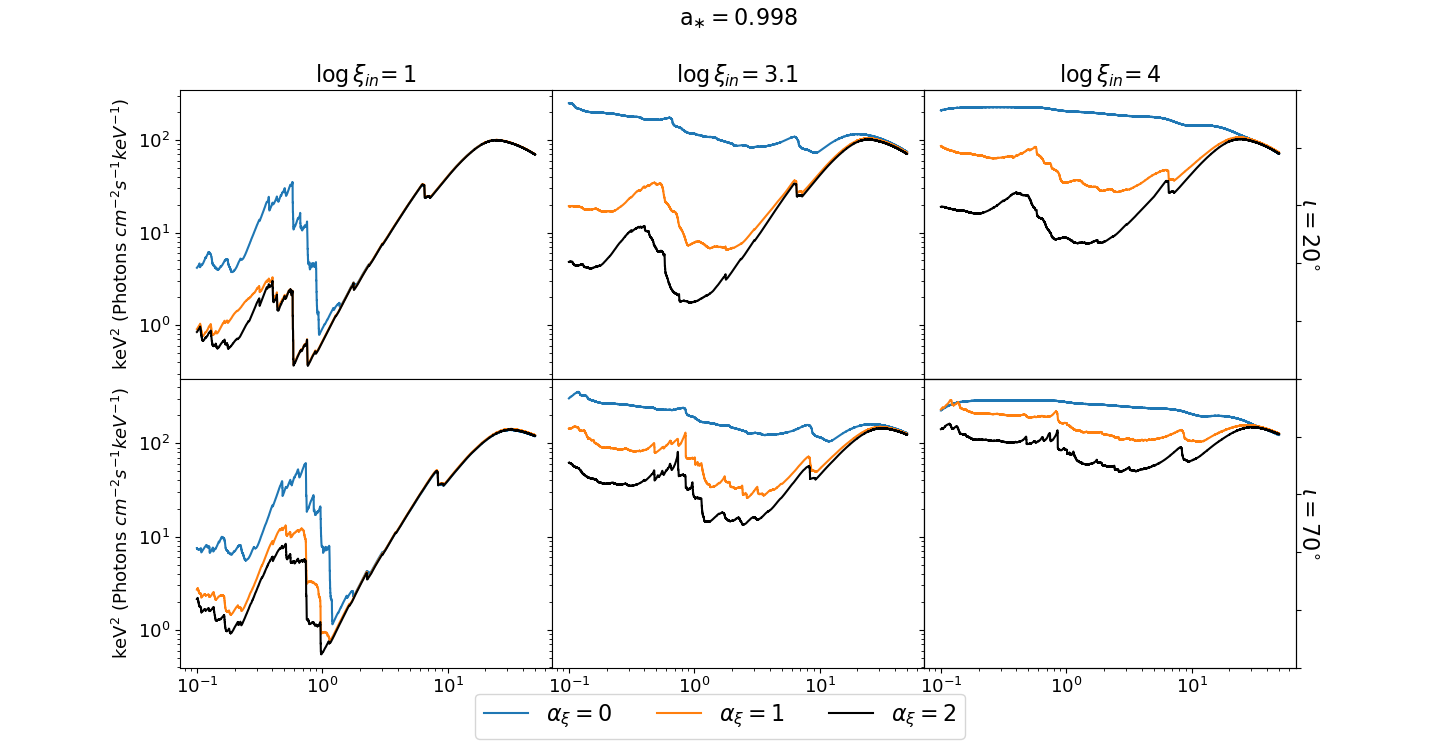}
    \caption{Synthetic reflection spectra calculated by {\tt relxillion\_nk} for $\log\xi_{\rm in} = 1$, 3.1, and 4, $\alpha_\xi = 0$, 1, and 2, and a disk's inclination angle $\iota = 20^\circ$ and $70^\circ$. All spectra are calculated in the Kerr spacetime with spin parameter $a_* = 0.998$ assuming that the inner edge of the disk $R_{\rm in}$ is at the innermost stable circular orbit, the spectrum illuminating the disk has photon index $\Gamma=2$ and high-energy cutoff $E_{\rm cut}=300$~keV, the emissivity profile of the disk is described by a power law with emissivity index $q=3$, and that the disk has Solar iron abundance, $A_{\rm Fe}=1$.}
    \label{fig:001}
\end{figure*}

\begin{figure*}
	\includegraphics[width=0.95\linewidth,trim=2.5cm 0.0cm 3.0cm 0.0cm,clip]{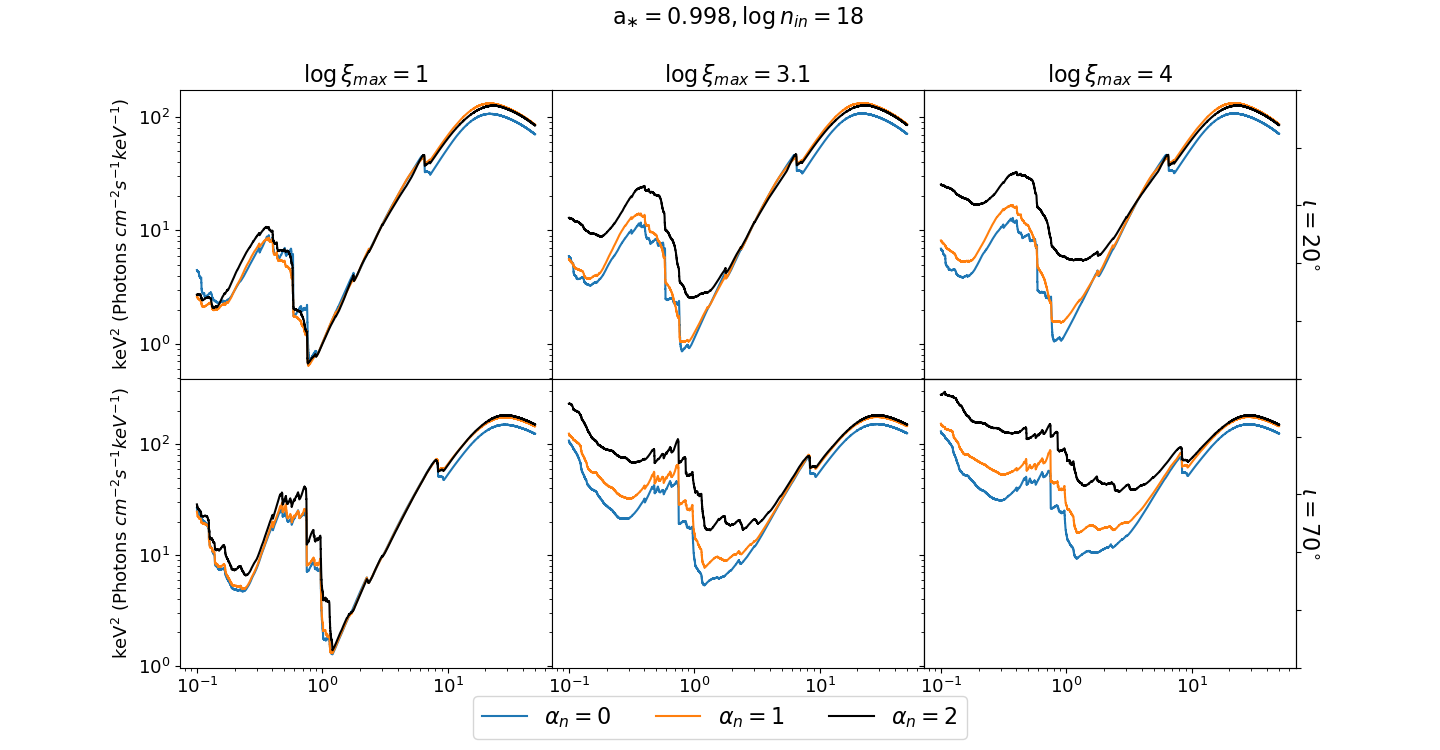}
    \caption{Synthetic reflection spectra calculated by {\tt relxilldgrad\_nk} for $\log\xi_{\rm max} = 1$, 3.1, and 4, $\alpha_n = 0$, 1, and 2, and a disk's inclination angle $\iota = 20^\circ$ and $70^\circ$. All spectra are calculated in the Kerr spacetime with spin parameter $a_* = 0.998$ assuming that the inner edge of the disk $R_{\rm in}$ is at the innermost stable circular orbit, the spectrum illuminating the disk has photon index $\Gamma=2$ and high-energy cutoff $E_{\rm cut}=300$~keV, the emissivity profile of the disk is described by a power law with emissivity index $q=3$, and that the disk has electron density at the inner edge of the accretion disk $\log n_{\rm in} = 19$ ($n_{\rm in}$ in units of cm$^{-3}$) and Solar iron abundance, $A_{\rm Fe}=1$.}
    \label{fig:002}
\end{figure*}

Even {\tt relxillion\_nk}~\citep{2021PhRvD.103j3023A} employs the {\tt xillver} table for the reflection spectrum in the rest-frame of the gas and therefore the electron density is assumed to be $10^{15}$~cm$^{-3}$ over the whole disk. The ionization has instead a radial profile described by a power law
\begin{eqnarray}\label{eq-xi-2}
\xi (r) = \xi_{\rm in} \left(\frac{R_{\rm in}}{r}\right)^{\alpha_\xi} \, .
\end{eqnarray}
The ionization of the disk is thus described by two parameters: the ionization parameter at the inner edge of the accretion disk $\xi_{\rm in}$ and the ionization index $\alpha_\xi$. $R_{\rm in}$ is the radial coordinate of the inner edge of the accretion disk, which is already a parameter of the model and in our analysis will be set at the innermost stable circular orbit. For $\alpha_\xi = 0$, {\tt relxillion\_nk} exactly reduces to {\tt relxill\_nk}, while for $\alpha_\xi > 0$ the value of the ionization parameter decreases as the radial coordinate $r$ increases. Even {\tt relxillion\_nk} is designed to calculate reflection spectra in non-Kerr spacetimes, but in this work we will not consider such a function and we will freeze the deformation parameter of the model to zero in order to impose the Kerr background. Fig.~\ref{fig:001} shows some synthetic reflection spectra calculated by {\tt relxillion\_nk} for different values of the ionization parameter at the inner edge of the accretion disk $\xi_{\rm in}$, ionization index $\alpha_\xi$, and inclination angle of the disk, assuming that the spacetime is described by the Kerr metric with spin parameter $a_* = 0.998$. For $\log\xi_{\rm in} = 1$ (left panels, $\xi_{\rm in}$ in units of erg~cm~s$^{-1}$), a non-vanishing ionization gradient only affects the spectrum below 2~keV, while for $\log\xi_{\rm in} = 3.1$ (central panels) and 4 (right panels) we see differences even at the iron line and Compton hump regions.

In {\tt relxilldgrad\_nk}~\citep{2021ApJ...923..175A}, the electron density profile is described by a power law and there are two parameters: the electron density at the inner edge of the accretion disk $n_{\rm in}$ and the electron density index $\alpha_n$
\begin{eqnarray}\label{eq-ne}
n_{\rm e} (r) = n_{\rm in} \left(\frac{R_{\rm in}}{r}\right)^{\alpha_n} \, .
\end{eqnarray}
Unlike {\tt relxill\_nk} and {\tt relxillion\_nk}, {\tt relxilldgrad\_nk} uses the table of {\tt xillverD} for the reflection spectrum in the rest-frame of the gas. The electron density $n_{\rm in}$ is thus allowed to vary in the range $10^{15}$~cm$^{-3}$ to $10^{19}$~cm$^{-3}$. However, in order to limit the size of the table, {\tt xillverD} assumes that the high-energy cutoff of the corona $E_{\rm cut}$ is fixed to 300~keV (in {\tt xillver}, $E_{\rm cut}$ can range from 5~keV to 1~MeV). The ionization parameter is calculated assuming that the X-ray flux from the corona $F_X$ is proportional to the emissivity profile of the reflection spectrum $\epsilon$\footnote{In the case of a point-like corona with a power law spectrum, we have $F_X \propto g^{2-\Gamma} \epsilon$, where $g = E_{\rm d}/E_{\rm c}$ is the redshift experienced by photons to travel from the corona to the disk and $\Gamma$ is the photon index of the power law spectrum of the corona. For $\Gamma = 2$, $F_X \propto \epsilon$, but this is not the case in general. However, the calculation of the redshift $g$ requires to know the coronal geometry (we have to know the locations of the emission point in the corona and of the absorption/scattering point on the disk). For an arbitrary coronal geometry, we can only make the approximation $F_X \propto \epsilon$. See \citet{2021ApJ...923..175A} for more details.} and we have
\begin{eqnarray}\label{eq-xi-3}
\xi (r) = \xi_{\rm max} \left[\frac{4 \pi \epsilon (r) }{n (r) }\right]_{\rm norm} \, ,
\end{eqnarray}
where $\xi_{\rm max}$ is the maximum value of the ionization parameter and the expression in square brackets is normalized with respect to such a maximum value. Note that $\xi_{\rm max}$ may not be at the inner edge of the accretion disk $R_{\rm in}$ in some systems, depending on the values of $n_{\rm in}$, $\alpha_n$, and $\epsilon (r)$. Even for {\tt relxilldgrad\_nk}, in this work we will set the deformation parameter to zero to work in the Kerr geometry. Fig.~\ref{fig:002} shows some synthetic reflection spectra calculated by {\tt relxilldgrad\_nk} for different values of the maximum ionization parameter $\xi_{\rm max}$, electron density index $\alpha_n$, and inclination angle of the disk, assuming that the spacetime is described by the Kerr metric with spin parameter $a_* = 0.998$ and the electron density at the inner edge of the accretion disk is $\log n_{\rm in} = 19$ ($n_{\rm in}$ in units of cm$^{-3}$). While the value of the electron density gradient mainly affects the spectrum below 3~keV, we have differences even in the iron line and Compton hump regions.


\section{Sources and observations}\label{sec:o}

For our study, we have selected three observations of three bright Galactic black holes. Sources and observations are listed in Tab.~\ref{tab:table1}

\begin{table*}
\caption{Summary of the sources and the observations analyzed in the present work.}
{\renewcommand{\arraystretch}{1.3}
	\begin{tabular}{l c c c c r}
	    \hline
		\hline
		Source & Satellite & Observation ID & Observation Date & Exposure (ks)\\
		\hline
        GS~1354--645 & \textsl{NuSTAR} & 90101006004 & 2015 July 11 & 30.0 \\\hline
        	GRS~1739--278 & \textsl{NuSTAR} & 80002018002 & 2014 March 26 & 29.7 \\\hline
        GRS~1915+105 & \textsl{Suzaku} & 402071010 & 2007 May 7 &  117 \\\hline
		\hline
	\end{tabular}
		}\label{tab:table1}
 \end{table*}

\subsection{{GS~1354--645}}\label{sec:GS 1354-645} 

GS~1354--645 was discovered with the All Sky Monitor aboard the \textsl{Ginga} satellite during its outburst in 1987~\citep{1990ApJ...361..590K}. The source went in an outburst again in 1997 and was observed with \textsl{RXTE}~\citep{2000ApJ...530..955R}. In June 2015, \textsl{Swift}/BAT detected a new outburst of GS~1354--645: interestingly, the source was only found in the low/hard spectral state~\citep{2015ATel.7612....1M}. \textsl{NuSTAR} observed the source in July. The \textsl{NuSTAR} data show very strong reflection features and the spectral analysis suggests a high inclination angle of the disk, a black hole spin parameter close to 1, and a very steep emissivity profile in the inner part of the accretion disk~\citep{El_Batal_2016}.

\subsection{{GRS~1739--278}}\label{sec:GRS 1739-278} 

GRS~1739--278 was discovered with the SIGMA telescope aboard \textsl{Granat}~\citep{1991AdSpR..11h.289P}. The X-ray spectral and timing characteristics strongly suggested that the compact object was a black hole. GRS~1739--278 went into outburst again, after an extended quiescent period, in March~2014 and was detected by \textsl{Swift}/BAT~\citep{2014ATel.5986....1K}. The source was observed even by \textsl{NuSTAR} and the analysis of the data was reported in \citet{Miller_2015}. The \textsl{NuSTAR} spectrum presents strong relativistic reflection features. \citet{Miller_2015} find a high black hole spin parameter, a steep inner emissivity profile, and a low inclination angle of the accretion disk.

\subsection{GRS~1915+105}\label{sec:GRS 1915+105} 

GRS~1915+105 is quite a peculiar source. It is a low-mass X-ray binary, as the mass of the companion star is less than 1~$M_\odot$. GRS~1915+105 was discovered by \textsl{Granat} in 1992~\citep{1992IAUC.5590....2C}, and since then it has never returned to the quiescent state. GRS~1915+105 shows remarkable variability across the whole electromagnetic spectrum \citep{2000A&A...355..271B} and is identified as a micro-quasar because of its radio jets~\citep{1999ARA&A..37..409M}. \citet{Zhang_1997} were the first to report a very high spin parameter of this black hole from the analysis of the thermal spectrum of the disk, later confirmed by \citet{McClintock_2006} using \textsl{RXTE} data and a more sophisticated theoretical model. The discovery of a relativistically broadened iron K$\alpha$ emission line was reported in \citet{2002A&A...387..215M} using \textsl{BeppoSAX} data. The very high value of the black hole spin parameter has been confirmed by studies of the reflection features using \textsl{Suzaku}~\citep{Blum_2009} and \textsl{NuSTAR} data \citep{Miller_2013}. In our study, we re-analyze the \textsl{Suzaku} observation, which is more suitable for accurate and precise measurements of the properties of the source~\citep[see, e.g.,][]{2019ApJ...875...41Z,2019ApJ...884..147Z}. In particular, the \textsl{NuSTAR} spectrum analyzed in \citet{Miller_2013} presents a non-relativistic reflection component and from its analysis we cannot get an accurate measurement of the properties of the source~\citep{2019ApJ...875...41Z}.


\section {Data Reduction}\label{sec:dr}

We note that the \textsl{NuSTAR} observations of GS~1354--645 and GRS~1739--278 and the \textsl{Suzaku} observation of GRS~1915+105 were already analyzed with an earlier version of our reflection model {\tt relxill\_nk} in \citet{Tripathi_2021} and \citet{2020MNRAS.498.3565T}, respectively. We follow \citet{Tripathi_2021}  for the data reduction of the \textsl{NuSTAR} spectra and \citet{2020MNRAS.498.3565T} for the data reduction of the \textsl{Suzaku} data, and here we only report the main steps.

\subsection{Reduction of \textsl{NuSTAR} data}\label{sec:NuStar}
 
The data from the \textsl{NuSTAR} detectors (FPMA and FPMB) were processed to get clean events using the script {\tt nupipeline} of \textsl{NuSTAR} data analysis Software NustarDAS~v2.0.0, which was a part of spectral analysis software HEASOFT v6.28. We used the latest calibration database CALDB v20200912. The source region was selected to get 90\% of its photons. A background region of the same size as the source region was selected far from the source to avoid inclusion of source photons. The redistribution matrix file (RMF) and ancillary response file (ARF) were created using the module {\tt nuproducts} of NuSTARDAS.

\subsection{Reduction of \textsl{Suzaku} data}

\textsl{Suzaku} observed GRS~1915+105 on 2007 May 7 (Obs. ID~402071010) for approximately 117~ks. The event files from the XIS1 were processed with {\tt aepipeline} to create a clean event file, using XIS CALDB version 20160616. The source region was selected by an annulus region due to photon pile up in the center of the detector~\citep{Blum_2009}. The background region was also selected by an annulus region. The XIS redistribution matrix file (RMF) and ancillary response file (ARF) were created respectively using the tools {\tt xisrmfgen} and {\tt xissimarfgen} available in
the HEASOFT version 6.24 data reduction package. HXD/PIN data were reduced similarly, employing {\tt aepipeline} and then {\tt hxdpinxbpi} using the latest CALDB version 20110915.

\begin{figure}
	\includegraphics[width=0.95\linewidth,trim=2.0cm 0.75cm 4.5cm 18.5cm,clip]{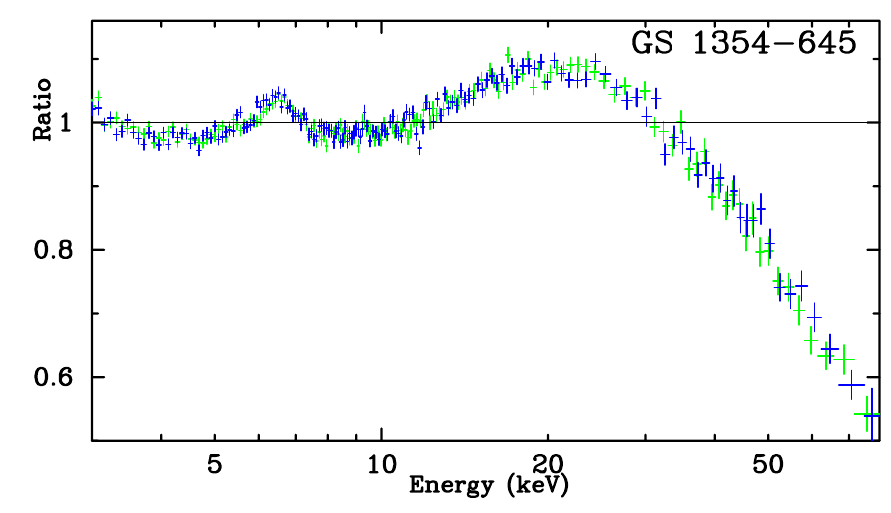}
    \caption{Data to best-fit model ratio for an absorbed power law for the \textsl{NuSTAR} spectrum of GS~1354--645 analyzed in this work. Blue and green crosses are for \textsl{NuSTAR}/FPMA and \textsl{NuSTAR}/FPMB data, respectively.}
    \label{fig:figure1a}
\end{figure}

\begin{figure}
	\includegraphics[width=0.95\linewidth,trim=2.0cm 0.75cm 4.5cm 18.5cm,clip]{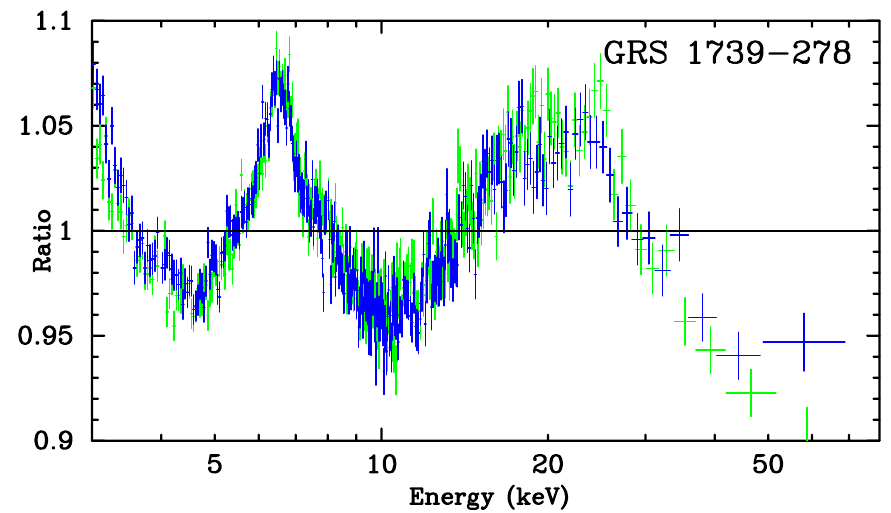}
    \caption{Data to best-fit model ratio for an absorbed power law for the \textsl{NuSTAR} spectrum of GRS~1739--278 analyzed in this work. Blue and green crosses are for \textsl{NuSTAR}/FPMA and \textsl{NuSTAR}/FPMB data, respectively.}
    \label{fig:figure1b}
\end{figure}

\begin{figure}
	\includegraphics[width=0.95\linewidth,trim=2.0cm 0.75cm 4.5cm 18.5cm,clip]{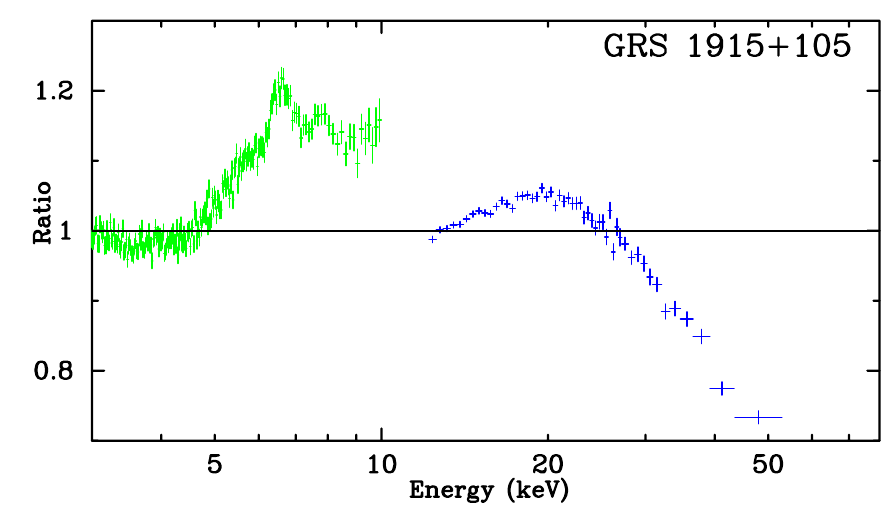}
    \caption{Data to best-fit model ratio for an absorbed power law for the \textsl{Suzaku} spectrum of GRS~1915+105 analyzed in this work. Green and blue crosses are for \textsl{Suzaku}/XIS and \textsl{Suzaku}/HXD data, respectively.}
    \label{fig:figure1c}
\end{figure}


\begin{table*}
\caption{Summary of the best-fit values from the analysis of the spectrum of GS~1354--645. $^*$ indicates that the parameter is frozen in the fit. The reported uncertainties correspond to the 90\% confidence level for one relevant parameter ($\Delta\chi^2 = 2.71$).}
{\renewcommand{\arraystretch}{1.3}
	\begin{tabular}{l c c c}
	    \hline
		\hline
Parameter & {\tt relxill\_nk} & {\tt relxillion\_nk} & {\tt relxilldgrad\_nk} \\\hline
{\tt tbabs} \\
$N_{\rm H}$ [$10^{22}$~cm$^{-2}$] & $0.7^*$ & $0.7^*$ & $0.7^*$\\\hline
{\tt cutoffpl} \\
$\Gamma$ & -- & -- & {1.692 $\substack{+0.022 \\ -0.018}$} \\
$E_{\rm cut}$ [keV] & -- & -- & {92 $\substack{+7 \\ -11}$} \\
norm & -- & -- & {0.336 $\substack{+0.023 \\ -0.003}$}\\\hline
{\tt relxill\_nk} \\
$q_{\rm in}$ & $10_{-0.21}$ & $10_{-0.6}$ & $10_{-5}$ \\
$q_{\rm out}$ & {0.90 $\substack{+0.09 \\ -0.05}$} & {1.00 $\substack{+0.22 \\ -0.16}$} & {0.80 $\substack{+0.09 \\ -0.09}$}\\
$R_{\rm br}$ [$r_{\rm g}$] & {4.1 $\substack{+0.8 \\ -0.4}$ } & {4.1 $\substack{+0.8 \\ -0.4}$} & {3.18 $\substack{+0.08 \\ -0.08}$} \\
$a_*$ & {0.993 $\substack{+0.002 \\ -0.003}$}  & {0.991 $\substack{+0.004 \\ -0.038}$} & $0.998_{-0.024}$ \\
$i$ [deg] & {76.6 $\substack{+1.1 \\ -3.2}$} & {73.8 $\substack{+2.0 \\ -6.2}$} & {81.9 $\substack{+0.5 \\ -2.4}$}  \\
$\Gamma'$ &  {1.626 $\substack{+0.006 \\ -0.003}$} & {1.53 $\substack{+0.02 \\ -0.02}$} & $=\Gamma$ \\
$\log\xi$ [erg cm $s^{-1}$] & {2.18 $\substack{+ 0.04 \\ -0.04}$} & -- & -- \\
$\log\xi_{\rm in}$ [erg cm $s^{-1}$] & -- & {3.60 $\substack{+0.13 \\ -0.13}$} & -- \\
$a_\xi$ & -- & {0.22 $\substack{+0.03 \\ -0.02 }$} & -- \\
$\log\xi_{\rm max}$ [erg cm $s^{-1}$] & -- & -- & {2.24 $\substack{+0.16 \\ -0.27}$} \\
$A_{\rm Fe}$ & {0.74 $\substack{+0.06 \\ -0.06}$} & {0.59 $\substack{+0.09 \\ -0.06}$} & {0.65 $\substack{+0.02 \\ -0.06}$} \\
$E'_{\rm cut}$ [keV] & {128 $\substack{+4 \\ -3}$} & {232 $\substack{+17 \\ -14}$} & $300^*$ \\
$\log n_{\rm e}$ [cm$^{-3}$] & $15^*$ & $15^*$ & -- \\
$\log n_{\rm in}$ [cm$^{-3}$] & -- & -- & {18.51 $\substack{+0.15 \\ -0.15}$}\\
$\alpha_{\rm n}$ & -- & -- & {7.52 $\substack{+0.26 \\ -1.15}$} \\
$R_{\rm f}$ & {0.269 $\substack{+0.019 \\ -0.013}$} & {1.60 $\substack{+0.96 \\ -0.51}$} & -- \\
norm [10$^{-3}$] & {8.881 $\substack{+0.016 \\ -0.079}$} & {2.6 $\substack{+1.0 \\ -0.6}$} & {3.00 $\substack{+0.05 \\ -0.13}$} \\\hline
$\chi^2/\nu$ & 2898.93/2724 & 2885.08/2723 & 2890.53/2722 \\
& =1.06422 & =1.05952 & =1.06191 \\
\hline
\hline
	\end{tabular}}
	\label{tab:table3}
\end{table*}

\begin{figure}
	\includegraphics[width=0.95\linewidth,trim=2.0cm 0.65cm 4.5cm 18.0cm,clip]{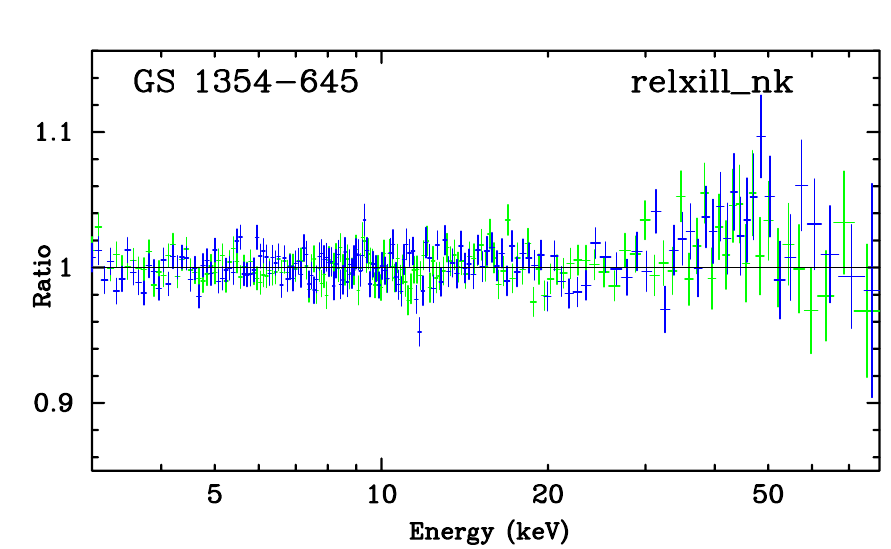} 
	\includegraphics[width=0.95\linewidth,trim=2.0cm 0.65cm 4.5cm 18.0cm,clip]{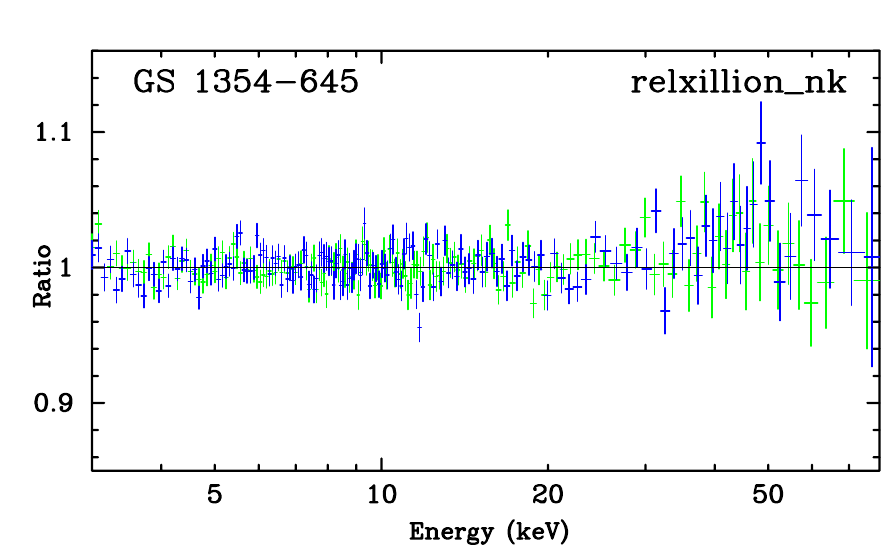}
	\includegraphics[width=0.95\linewidth,trim=2.0cm 0.65cm 4.5cm 18.0cm,clip]{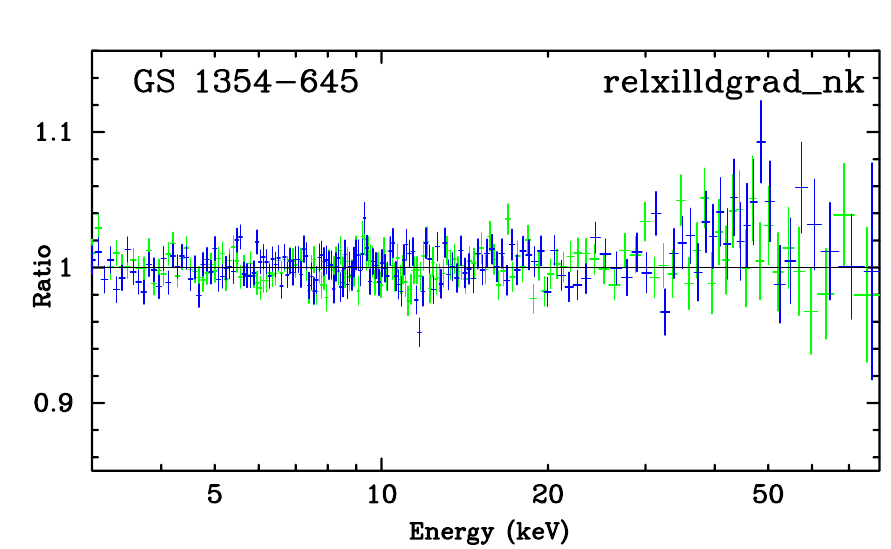}
    \caption{Data to best-fit model ratios for GS~1354--645 when we use {\tt relxill\_nk}, {\tt relxillion\_nk}, and {\tt relxilldgrad\_nk}. Blue and green crosses are for \textsl{NuSTAR}/FPMA and \textsl{NuSTAR}/FPMB data, respectively.}
    \label{fig:figure2}
\end{figure}

\begin{table*}
\caption{Summary of the best-fit values from the analysis of the spectrum of GRS~1739--278. $^*$ indicates that the parameter is frozen in the fit. The reported uncertainties correspond to the 90\% confidence level for one relevant parameter ($\Delta\chi^2 = 2.71$).}
{\renewcommand{\arraystretch}{1.3}
	\begin{tabular}{l c c c}
	    \hline
		\hline
Parameter & {\tt relxill\_nk} & {\tt relxillion\_nk} & {\tt relxilldgrad\_nk} \\\hline	
{\tt tbabs} \\
$N_{\rm H}$ [$10^{22}$ $cm^{-2}$] & 1.58 $\substack{+0.15 \\ -0.17}$ & 1.04 $\substack{+0.23 \\ -0.21}$ & 2.99 $\substack{+0.19 \\ -0.04}$ \\\hline
{\tt cutoffpl} \\
$\Gamma$ & -- & -- & {1.586 $\substack{+0.026 \\ -0.007}$} \\
$E_{\rm cut}$ [keV] & -- & -- & {17.2 $\substack{+0.6 \\ -0.8}$} \\
norm & -- & -- & {0.406 $\substack{+0.004 \\ -0.014}$}\\\hline
{\tt relxill\_nk} \\
$q_{\rm in}$ & 6.7 $\substack{+0.6 \\ -0.5}$ & $10_{-2.5}$ & 7.56 $\substack{+0.02 \\ -0.08}$ \\
$q_{\rm out}$ & {2.12 $\substack{+0.11 \\ -0.09}$} & {2.73 $\substack{+0.08 \\ -0.10}$} & {3.09 $\substack{+0.09 \\ -0.14}$} \\
=$R_{\rm br}$ [$r_{\rm g}$] & {5.5 $\substack{+0.8 \\ -0.8}$ } & {3.55 $\substack{+0.65 \\ -0.24}$} & {4.52 $\substack{+0.09 \\ -0.04}$} \\
$a_*$ & {0.977 $\substack{+0.017 \\ -0.017}$}  & {0.946 $\substack{+0.015 \\ -0.026}$} & {0.975 $\substack{+0.005 \\ -0.002}$} \\
$i$ [deg] & {16 $\substack{+5 \\ -12}$} & {23 $\substack{+4 \\-4}$} & $4^{+12}$ \\
$\Gamma'$ &  {1.201 $\substack{+0.027 \\ -0.017}$} & {1.18 $\substack{+0.03 \\ -0.04}$} & $=\Gamma$ \\
$\log\xi$ [erg cm $s^{-1}$] & {3.48 $\substack{+ 0.06 \\ -0.05}$} & -- & -- \\
$\log\xi_{\rm in}$ [erg cm $s^{-1}$] & -- & {4.37 $\substack{+0.10 \\ -0.10}$} & -- \\
$a_\xi$ & -- & {0.33 $\substack{+0.04 \\ -0.03 }$} & -- \\
$\log\xi_{\rm max}$ [erg cm $s^{-1}$] & -- & -- & {3.687 $\substack{+0.008 \\ -0.022}$}\\
$A_{\rm Fe}$ & {3.3 $\substack{+0.6 \\ -0.4}$} & {4.1 $\substack{+1.1 \\ -1.1}$} & {1.79 $\substack{+0.08 \\ -0.18}$} \\
$E'_{\rm cut}$ [keV] & {25.4 $\substack{+0.9 \\ -0.7}$} & {43.8 $\substack{+1.9 \\ -2.2}$} & $300^*$ \\
$\log n_{\rm e}$ [cm$^{-3}$] & $15^*$ & $15^*$ & -- \\
$\log n_{\rm in}$ [cm$^{-3}$] & -- & -- & {17.15 $\substack{+0.09 \\ -0.09}$}\\
$\alpha_{\rm n}$ & -- & -- & {7.908  $\substack{+0.149 \\ -0.006}$} \\
$R_{\rm f}$ & {0.42 $\substack{+0.07 \\ -0.07}$} & {2.4 $\substack{+0.6 \\ -0.3}$} & -- \\
norm [10$^{-3}$] & {9.2 $\substack{+0.5 \\ -0.5}$} & {3.0 $\substack{+0.3 \\ -0.5}$} & {12.4 $\substack{+2.8 \\ -2.6}$} \\\hline
{\tt xillver} \\
norm [10$^{-3}$] & -- & -- & 3.20 $\substack{+0.06 \\ -0.15}$ \\\hline
$\chi^2/\nu$ & 1344.41/1122 & 1298.27/1121 & 1334.66/1119\\
& =1.19823 & =1.15814 & =1.19273 \\
\hline
\hline
	\end{tabular}}
	\label{tab:table4}
\end{table*}

\begin{figure}
	\includegraphics[width=0.95\linewidth,trim=2.0cm 0.65cm 4.5cm 18.0cm,clip]{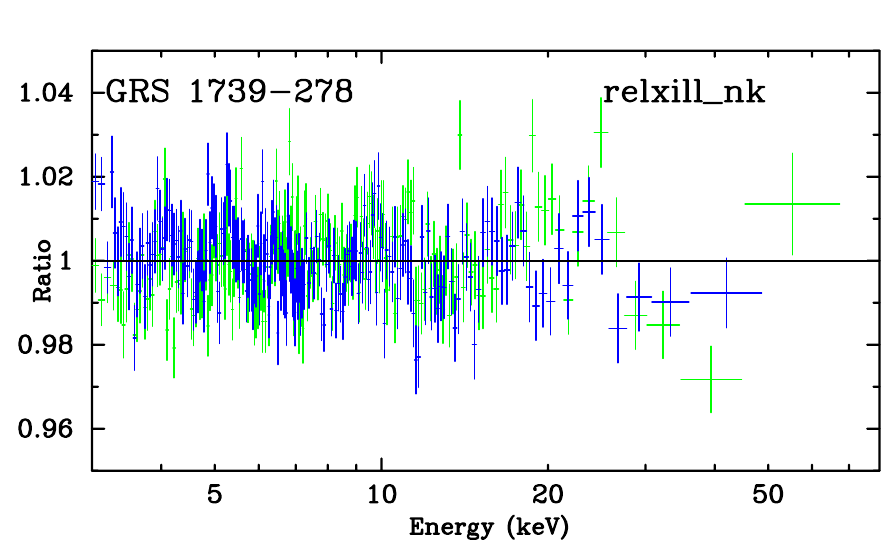}
	\includegraphics[width=0.95\linewidth,trim=2.0cm 0.65cm 4.5cm 18.0cm,clip]{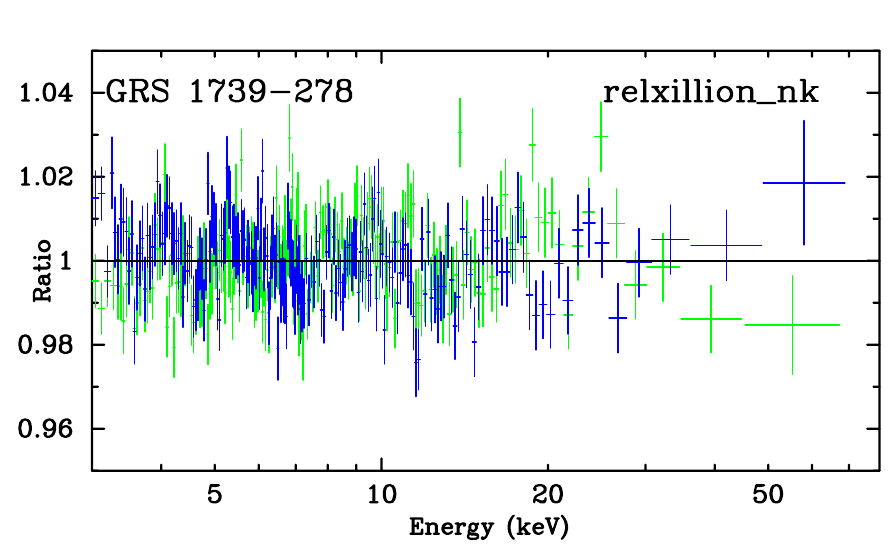}
	\includegraphics[width=0.95\linewidth,trim=2.0cm 0.65cm 4.5cm 18.0cm,clip]{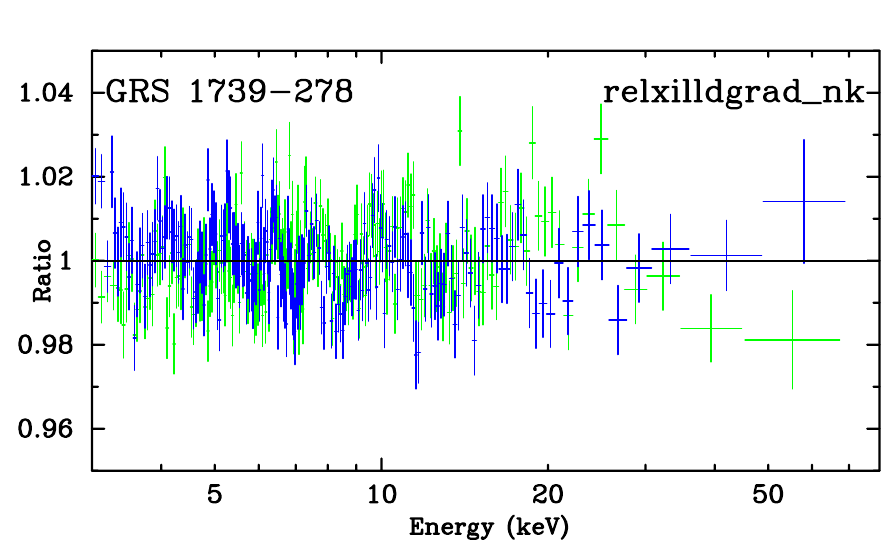}
    \caption{Data to best-fit model ratios for GRS~1739--278 when we use {\tt relxill\_nk}, {\tt relxillion\_nk}, and {\tt relxilldgrad\_nk}. Blue and green crosses are for \textsl{NuSTAR}/FPMA and \textsl{NuSTAR}/FPMB data, respectively.}
    \label{fig:figure3}
\end{figure}

\begin{table*}
\caption{Summary of the best-fit values from the analysis of the spectrum of GRS~1915+105. $^*$ indicates that the parameter is frozen in the fit. The reported uncertainties correspond to the 90\% confidence level for one relevant parameter ($\Delta\chi^2 = 2.71$). $^\dag$ indicates that the parameter cannot be constrained.}
{\renewcommand{\arraystretch}{1.3}
	\begin{tabular}{l c c c}
	    \hline
		\hline
Parameter & {\tt relxill\_nk} & {\tt relxillion\_nk} & {\tt relxilldgrad\_nk} \\\hline
{\tt tbabs} \\
$N_{\rm H}$ [$10^{22}$ $cm^{-2}$] & 7.57 $\substack{+0.03 \\ -0.09}$ & 7.86 $\substack{+0.03 \\ -0.14}$ & 8.35 $\substack{+0.07 \\ -0.07}$ \\\hline
{\tt cutoffpl} \\
$\Gamma$ & -- & -- & {2.605 $\substack{+0.022 \\ -0.008}$} \\
$E_{\rm cut}$ [keV] & -- & -- & $392^\dag$ \\
norm & -- & -- & {7.251 $\substack{+0.215 \\ -0.092}$}\\\hline
{\tt relxill\_nk} \\
$q_{\rm in}$ & 8.14 $\substack{+0.51 \\ -0.35}$ & 8.78 $\substack{+0.30 \\ -2.16}$ & $10_{-0.82}$ \\
$q_{\rm out}$ & $0^{+0.10}$ & $0^{+0.16}$ & {1.78 $\substack{+0.08 \\ -0.04}$} \\
$R_{\rm br}$ [$r_{\rm g}$] & {8.95 $\substack{+3.99 \\ -0.33}$ } & {7.75 $\substack{+0.55 \\ -0.31}$} & {2.33 $\substack{+0.09 \\ -0.14}$} \\
$a_*$ & {0.979 $\substack{+0.006 \\ -0.004}$}  & {0.985 $\substack{+0.004 \\ -0.007}$} & {0.996 $\substack{+0.001 \\ -0.002}$} \\
\rule{0pt}{11pt} $i$ [deg] & {67.5 $\substack{+0.5 \\ -1.8}$} & {69.5 $\substack{+0.4 \\ -1.5}$} & {79.4 $\substack{+3.2 \\ -0.9}$} \\
$\Gamma'$ &  {2.079 $\substack{+0.013 \\ -0.017}$} & {2.14 $\substack{+0.02 \\ -0.04}$} & $=\Gamma$ \\
$\log\xi$ [erg cm $s^{-1}$] & {3.07 $\substack{+ 0.02 \\ -0.05}$} & -- & -- \\
$\log\xi_{\rm in}$ [erg cm $s^{-1}$] & -- & {3.07 $\substack{+0.04 \\ -0.06}$} & -- \\
$a_\xi$ & -- & {0.09 $\substack{+0.03 \\ -0.02 }$} & -- \\
$\log\xi_{\rm max}$ [erg cm $s^{-1}$] & -- & -- & {2.50 $\substack{+0.07 \\ -0.01}$}\\
$A_{\rm Fe}$ & $0.5^{+0.03}$ & $0.5^{+0.02}$ & {1.22 $\substack{+0.14 \\ -0.14}$} \\
$E'_{\rm cut}$ [keV] & {59 $\substack{+2 \\ -1}$} & {64 $\substack{+2 \\ -3}$} & $300^*$ \\
$\log n_{\rm e}$ [cm$^{-3}$] & $15^*$ & $15^*$ & -- \\
$\log n_{\rm in}$ [cm$^{-3}$] & -- & -- & {$19_{-0.15}$}\\
$\alpha_{\rm n}$ & -- & -- & $0.001^{+0.054}$\\
$R_{\rm f}$ & {0.38 $\substack{+0.04\\ -0.03}$} & {0.46 $\substack{+0.02 \\ -0.61}$} & -- \\
norm [10$^{-2}$] & {3.6 $\substack{+0.2 \\ -0.1}$} & {3.9 $\substack{+0.1 \\ -0.2}$} & {8.5 $\substack{+0.2 \\ -0.4}$} \\\hline
$\chi^2/\nu$ & 2342.97/2209 & 2338.61/2208 & 2300.77/2207\\
& = 1.06065 & =1.05915 & =1.04249  \\
\hline
\hline
	\end{tabular}}
	\label{tab:table5}
\end{table*}

\begin{figure}
	\includegraphics[width=0.95\linewidth,trim=2.0cm 0.65cm 4.5cm 18.0cm,clip]{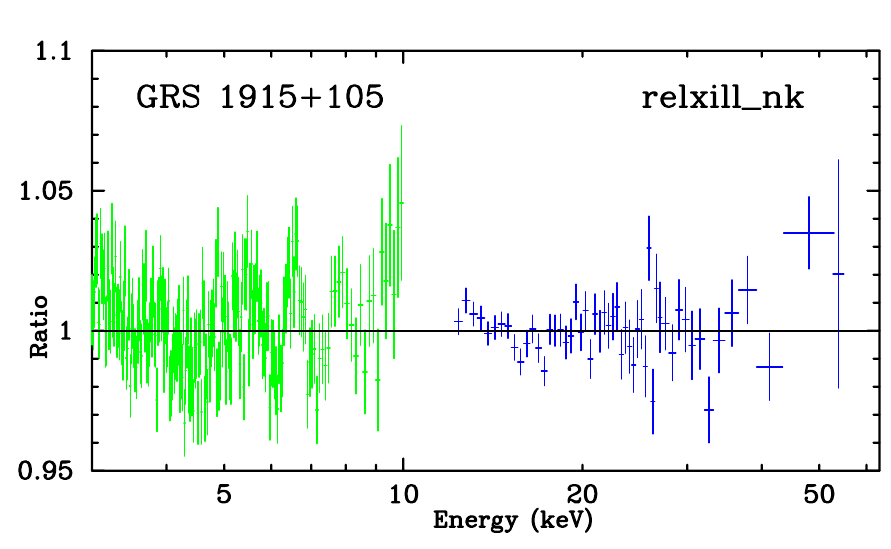}
	\includegraphics[width=0.95\linewidth,trim=2.0cm 0.65cm 4.5cm 18.0cm,clip]{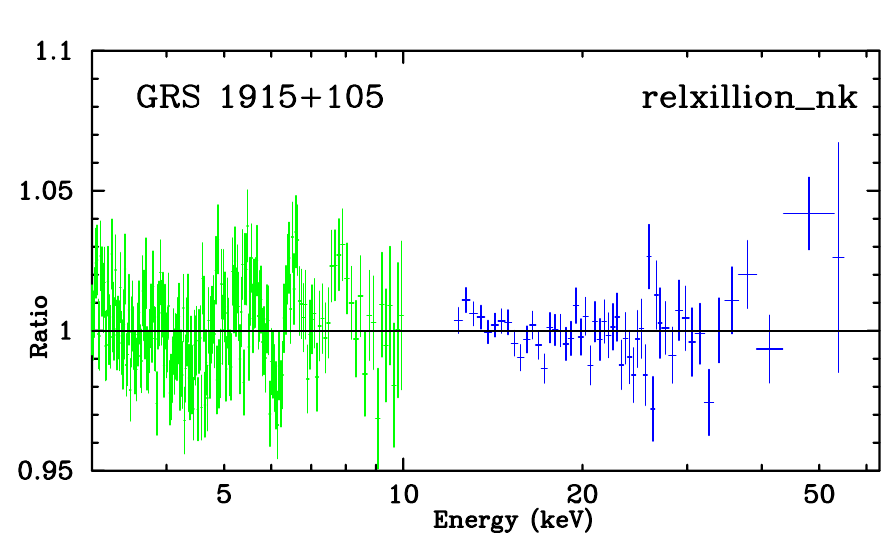}
	\includegraphics[width=0.95\linewidth,trim=2.0cm 0.65cm 4.5cm 18.0cm,clip]{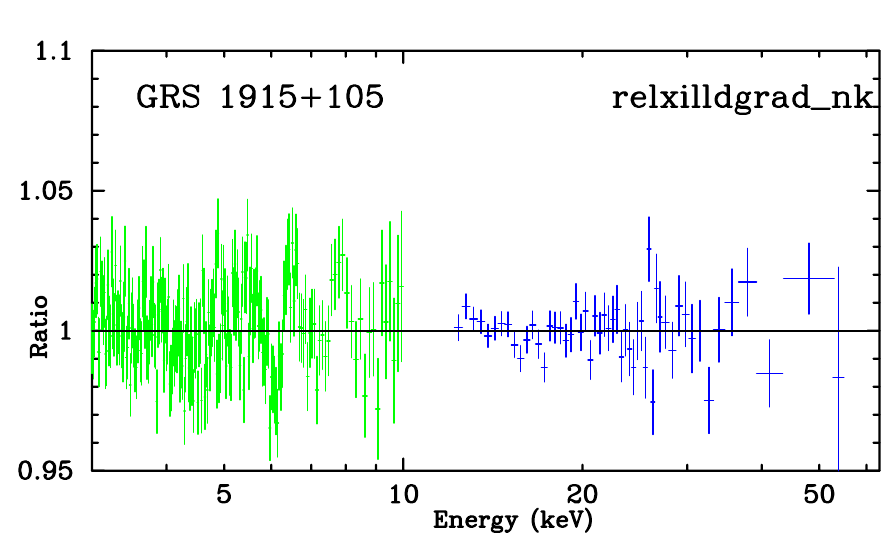}
    \caption{Data to best-fit model ratios for GRS~1915+105 when we use {\tt relxill\_nk}, {\tt relxillion\_nk}, and {\tt relxilldgrad\_nk}. Green and blue crosses are for \textsl{Suzaku}/XIS and \textsl{Suzaku}/HXD data, respectively.}
    \label{fig:figure4}
\end{figure}

\begin{figure}
	\includegraphics[width=0.95\linewidth,trim=0.0cm 0.0cm 0.0cm 0.0cm,clip]{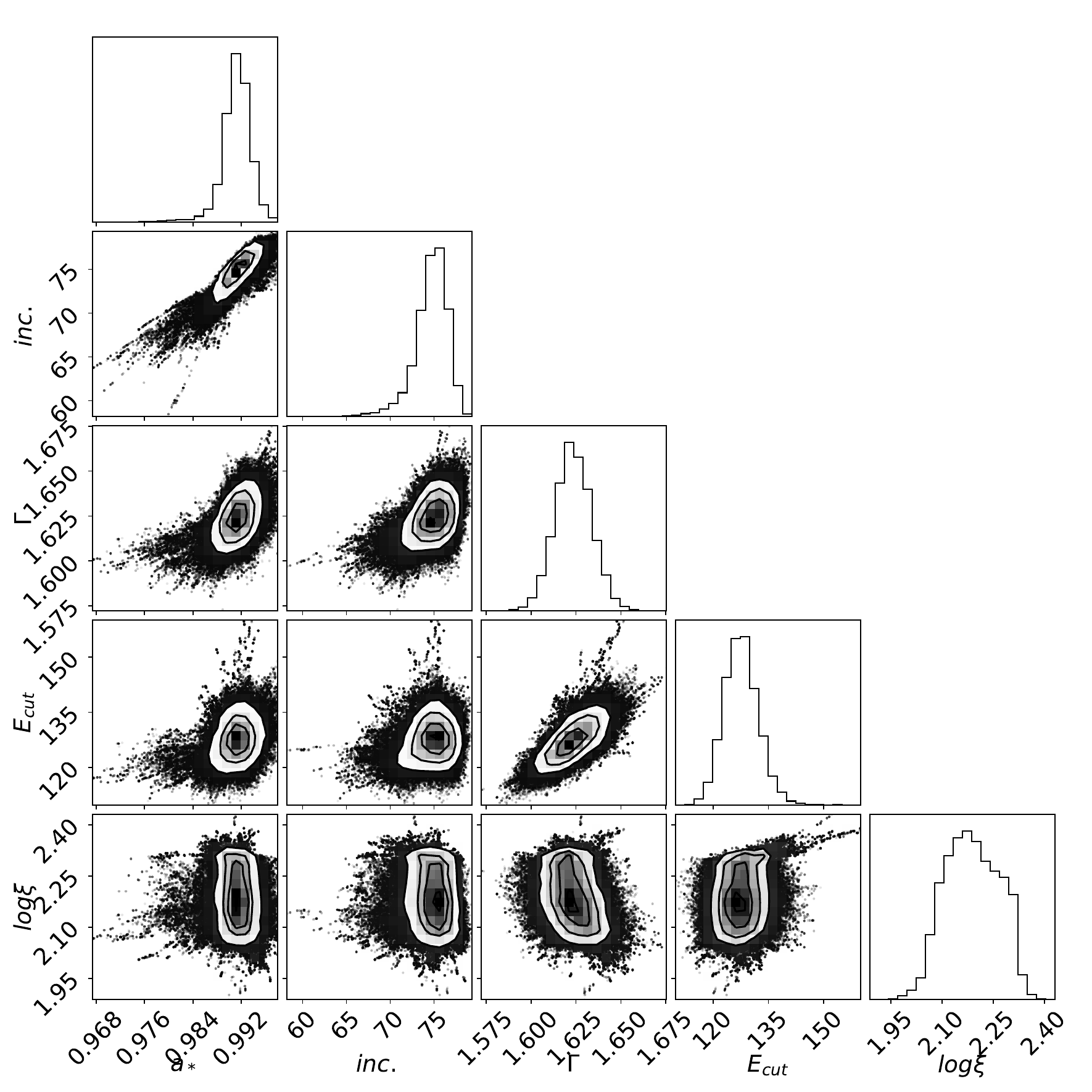}
    \caption{Corner plot for the key-parameter pairs (spin parameter $a_*$, inclination angle of the disk, photon index $\Gamma$, high-energy cutoff $E_{\rm cut}$, and ionization $\log\xi$) of the {\tt relxill\_nk} fit of GS~1354--645 after the MCMC run. The 2D plots report the 1, 2, and 3$\sigma$ confidence contours.}
    \label{fig:c1a}
\end{figure}

\begin{figure}
	\includegraphics[width=0.95\linewidth,trim=0.0cm 0.0cm 0.0cm 0.0cm,clip]{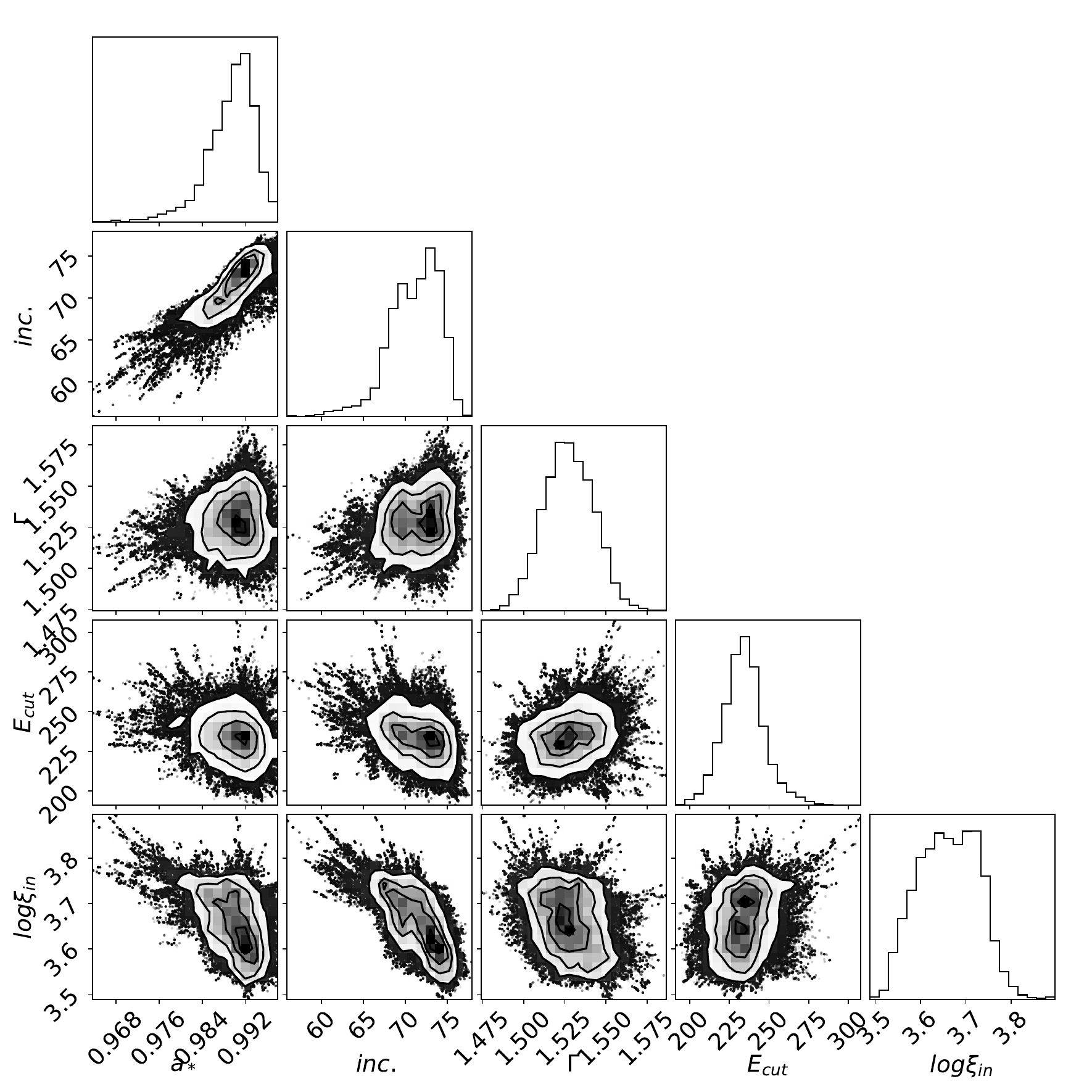}
    \caption{Corner plot for the key-parameter pairs (spin parameter $a_*$, inclination angle of the disk, photon index $\Gamma$, high-energy cutoff $E_{\rm cut}$, and ionization $\log\xi_{\rm in}$) of the {\tt relxillion\_nk} fit of GS~1354--645 after the MCMC run. The 2D plots report the 1, 2, and 3$\sigma$ confidence contours.}
    \label{fig:c1b}
\end{figure}

\begin{figure}
	\includegraphics[width=0.95\linewidth,trim=0.0cm 0.0cm 0.0cm 0.0cm,clip]{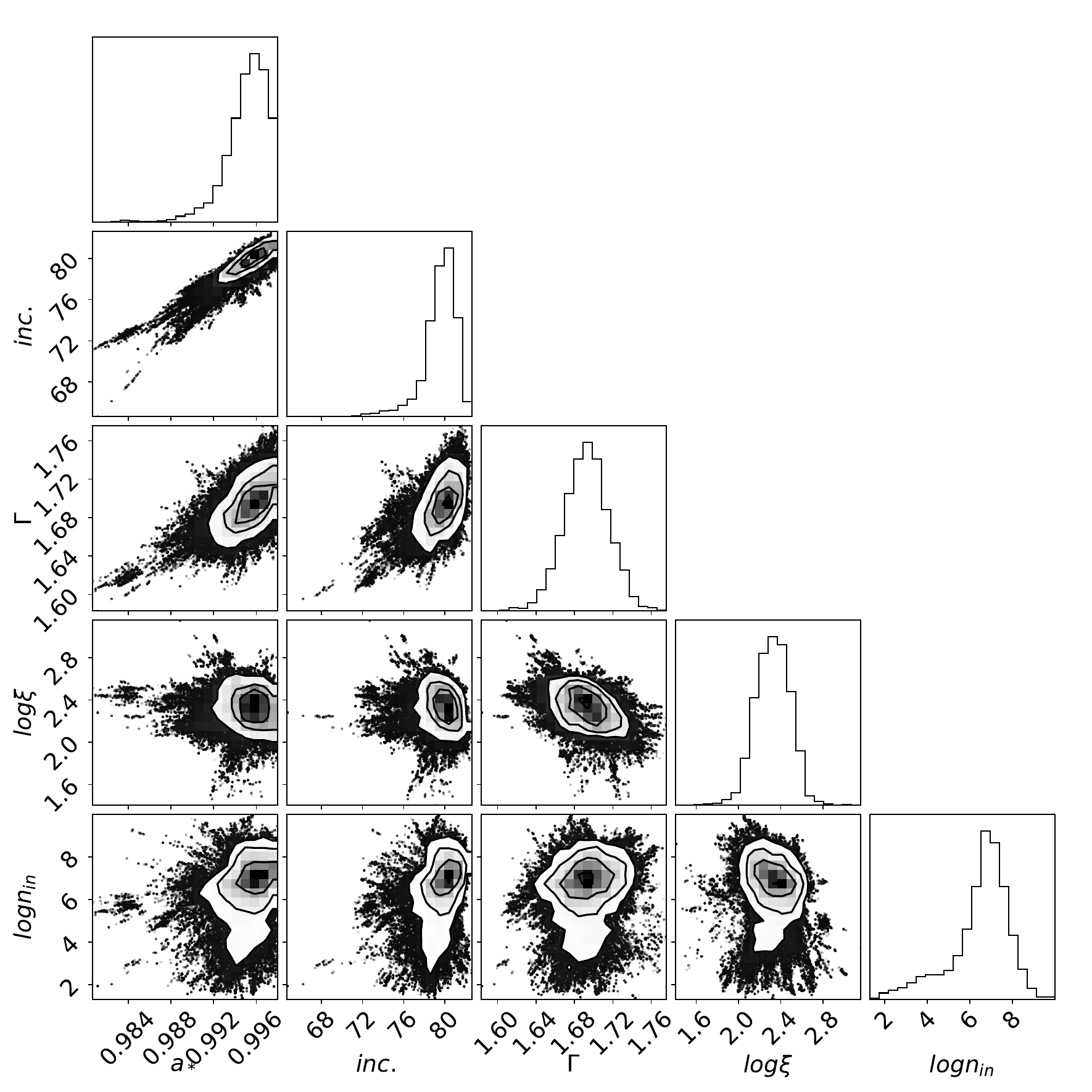}
    \caption{Corner plot for the key-parameter pairs (spin parameter $a_*$, inclination angle of the disk, photon index $\Gamma$, ionization $\log\xi$, and electron density $\log n_{\rm in}$) of the {\tt relxilldgrad\_nk} fit of GS~1354--645 after the MCMC run. The 2D plots report the 1, 2, and 3$\sigma$ confidence contours.}
    \label{fig:c1c}
\end{figure}

\begin{figure}
	\includegraphics[width=0.95\linewidth,trim=0.0cm 0.0cm 0.0cm 0.0cm,clip]{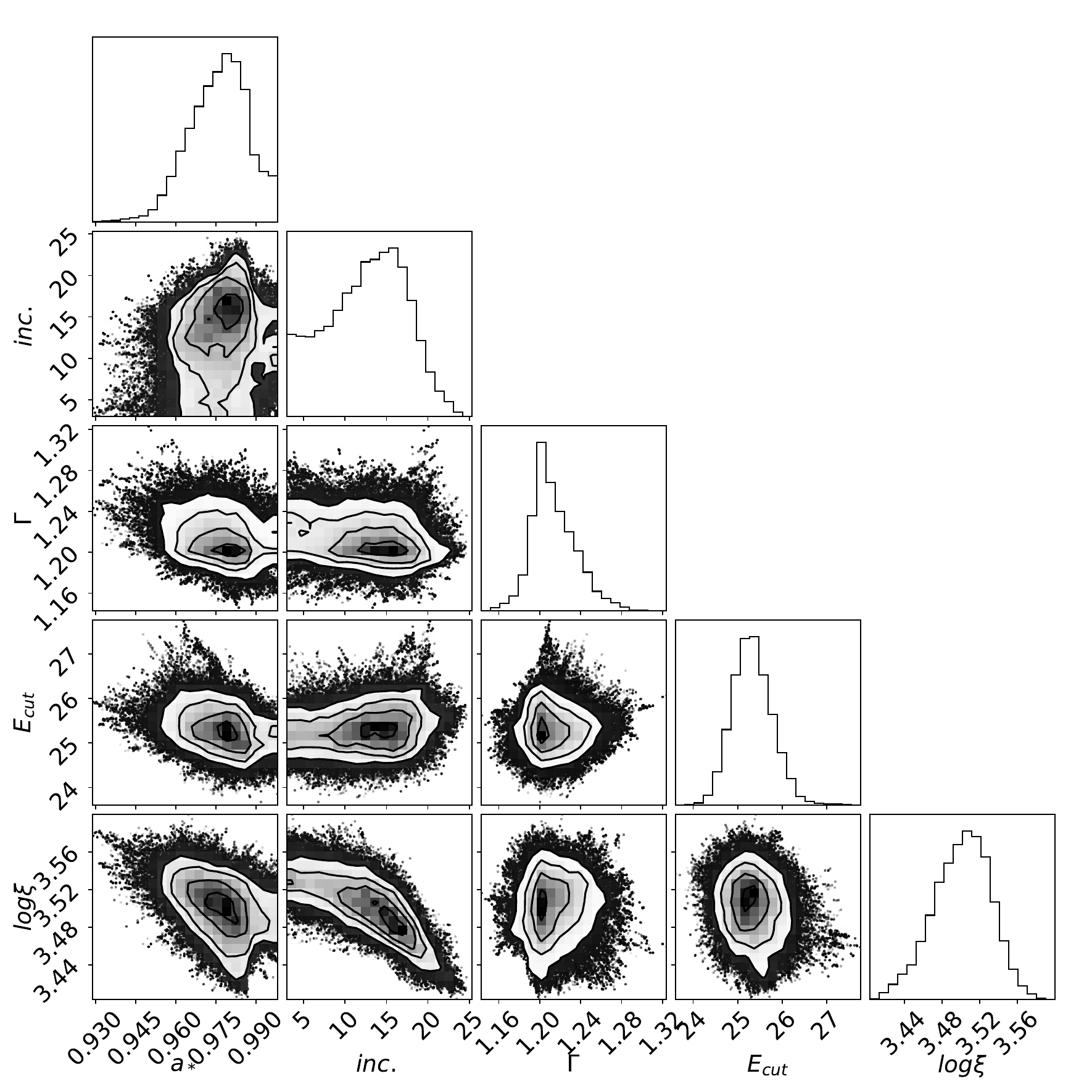}
    \caption{Corner plot for the key-parameter pairs (spin parameter $a_*$, inclination angle of the disk, photon index $\Gamma$, high-energy cutoff $E_{\rm cut}$, and ionization $\log\xi$) of the {\tt relxill\_nk} fit of GRS~1739--278 after the MCMC run. The 2D plots report the 1, 2, and 3$\sigma$ confidence contours.}
    \label{fig:c2a}
\end{figure}

\begin{figure}
	\includegraphics[width=0.95\linewidth,trim=0.0cm 0.0cm 0.0cm 0.0cm,clip]{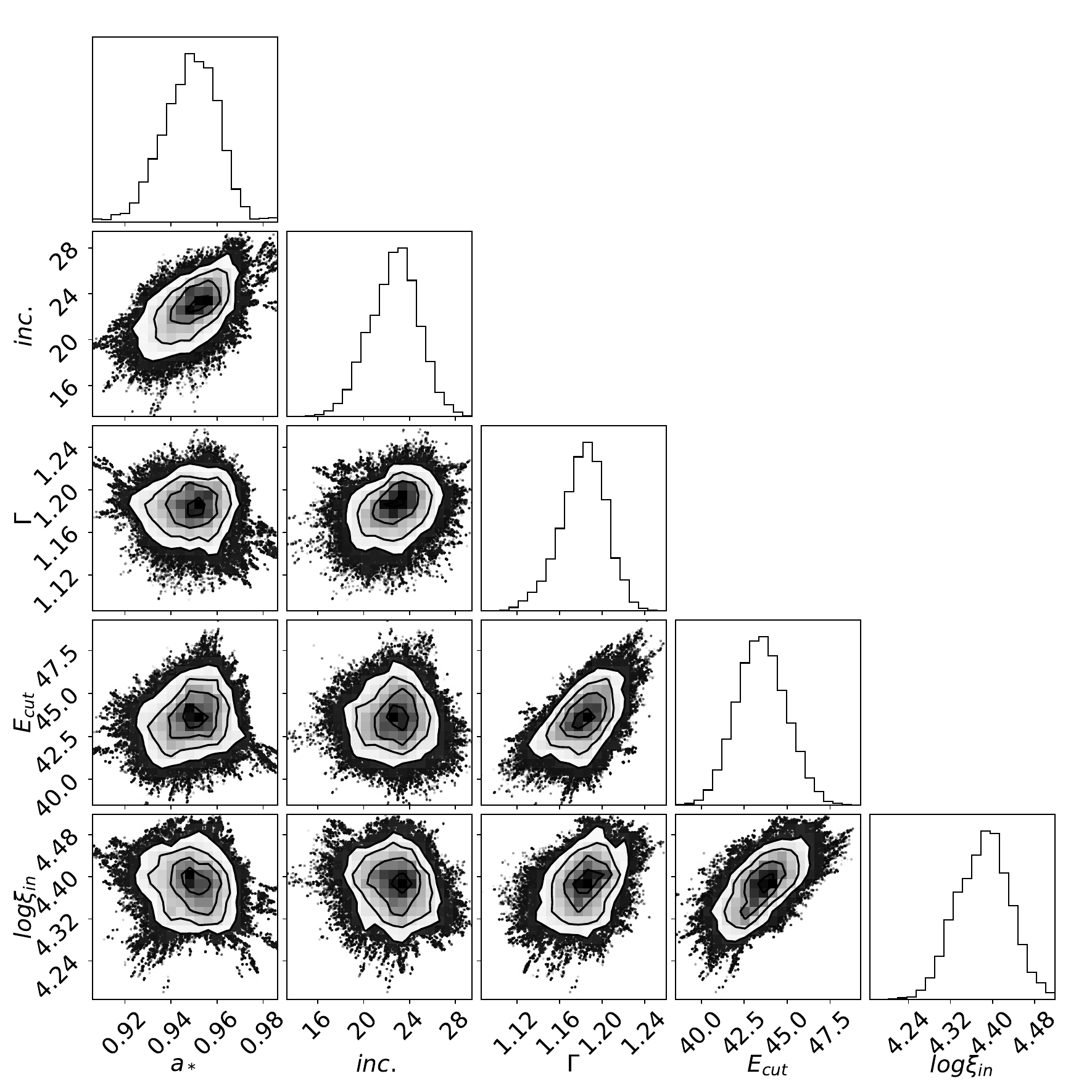}
    \caption{Corner plot for the key-parameter pairs (spin parameter $a_*$, inclination angle of the disk, photon index $\Gamma$, high-energy cutoff $E_{\rm cut}$, and ionization $\log\xi_{\rm in}$) of the {\tt relxillion\_nk} fit of GRS~1739--278 after the MCMC run. The 2D plots report the 1, 2, and 3$\sigma$ confidence contours.}
    \label{fig:c2b}
\end{figure}

\begin{figure}
	\includegraphics[width=0.95\linewidth,trim=0.0cm 0.0cm 0.0cm 0.0cm,clip]{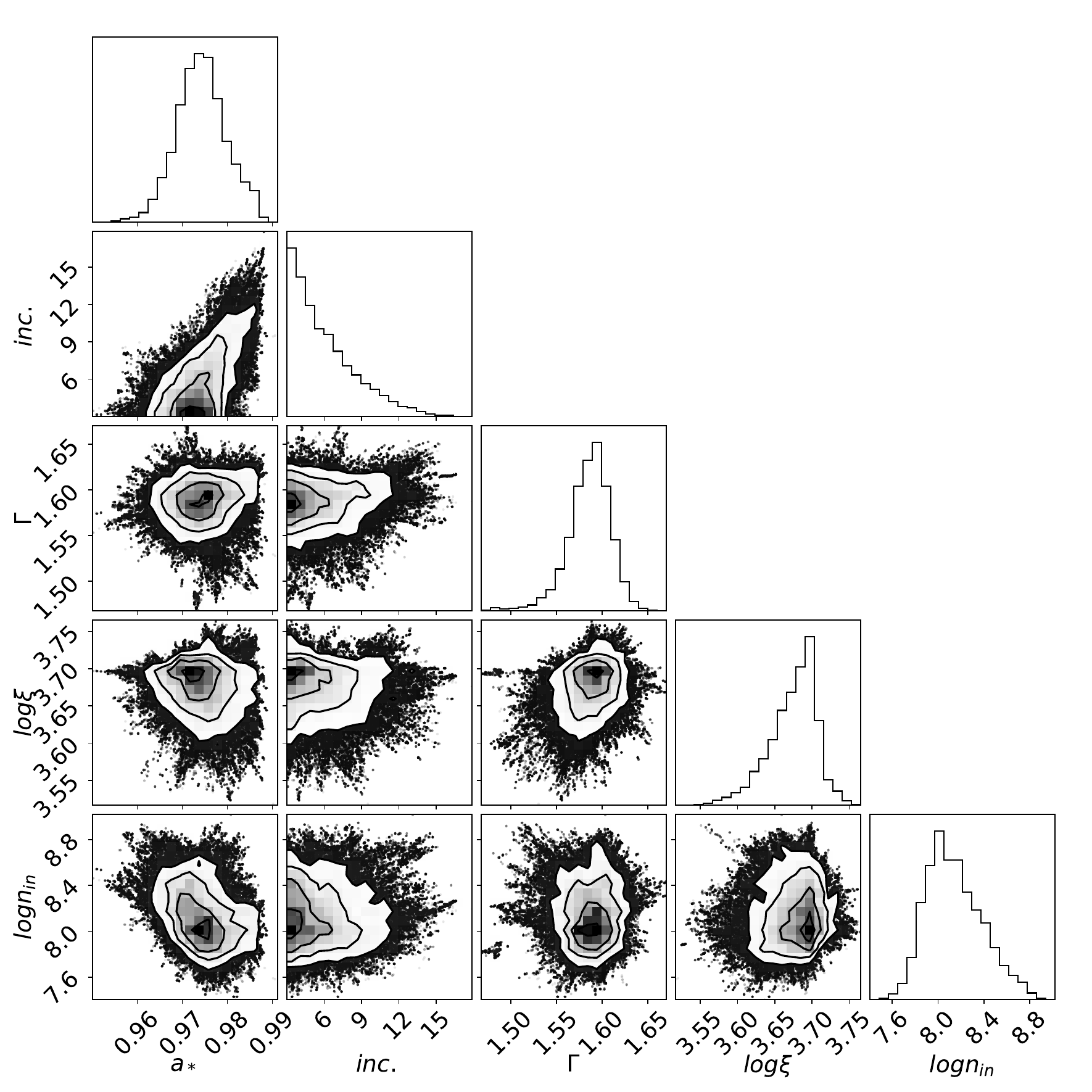}
    \caption{Corner plot for the key-parameter pairs (spin parameter $a_*$, inclination angle of the disk, photon index $\Gamma$, ionization $\log\xi$, and electron density $\log n_{\rm in}$) of the {\tt relxilldgrad\_nk} fit of GRS~1739--278 after the MCMC run. The 2D plots report the 1, 2, and 3$\sigma$ confidence contours.}
    \label{fig:c2c}
\end{figure}

\begin{figure}
	\includegraphics[width=0.95\linewidth,trim=0.0cm 0.0cm 0.0cm 0.0cm,clip]{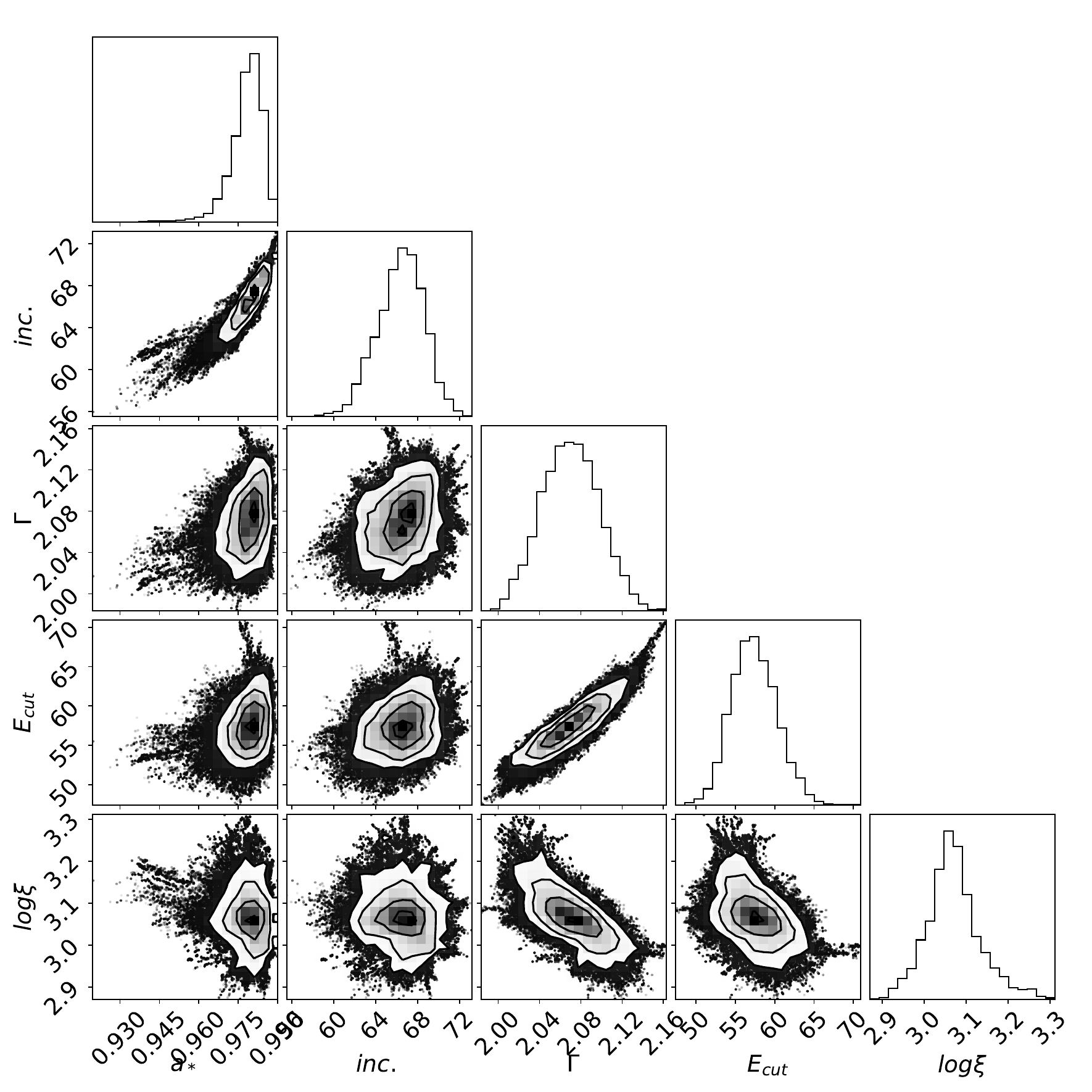}
    \caption{Corner plot for the key-parameter pairs (spin parameter $a_*$, inclination angle of the disk, photon index $\Gamma$, high-energy cutoff $E_{\rm cut}$, and ionization $\log\xi$) of the {\tt relxill\_nk} fit of GRS~1915+105 after the MCMC run. The 2D plots report the 1, 2, and 3$\sigma$ confidence contours.}
    \label{fig:c3a}
\end{figure}

\begin{figure}
	\includegraphics[width=0.95\linewidth,trim=0.0cm 0.0cm 0.0cm 0.0cm,clip]{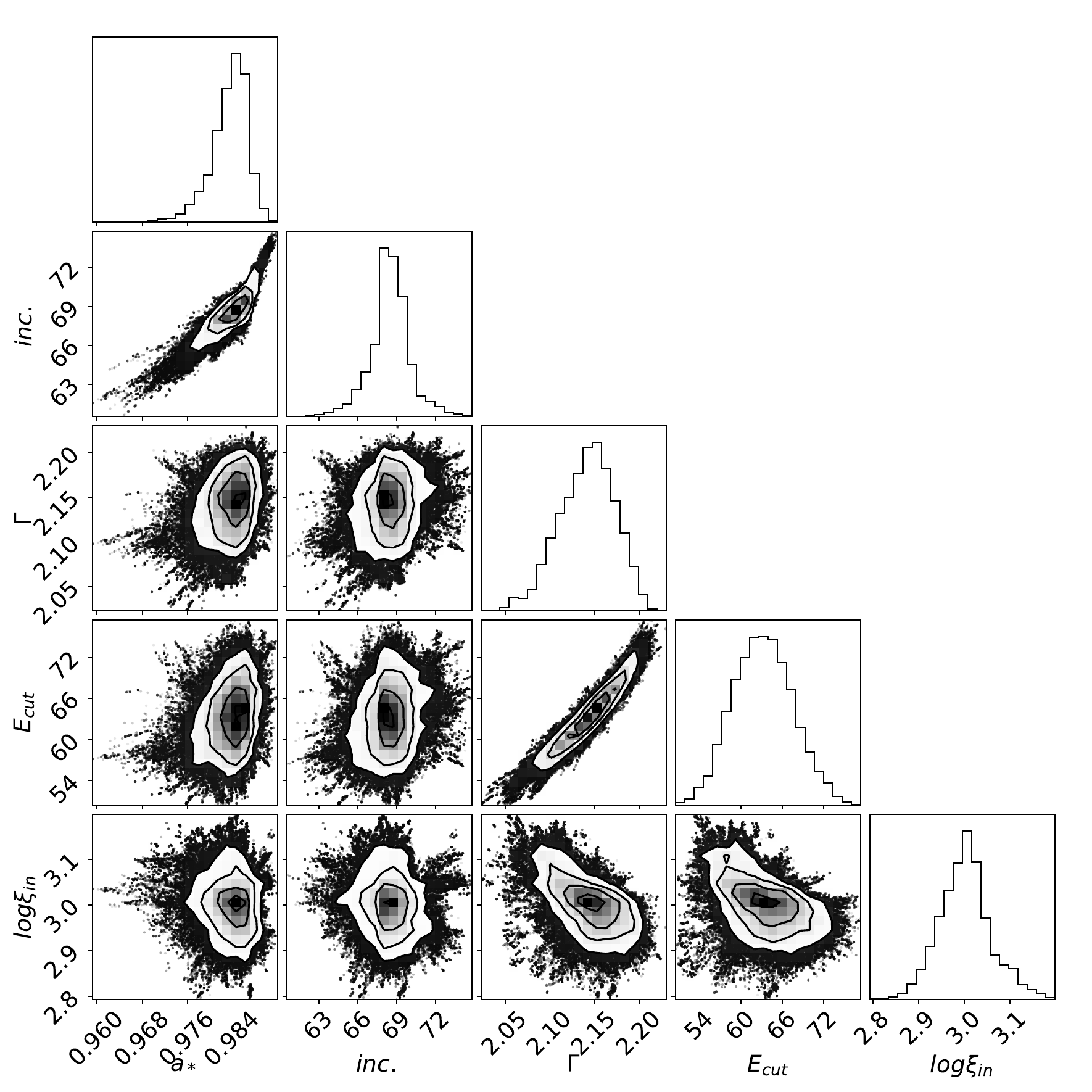}
    \caption{Corner plot for the key-parameter pairs (spin parameter $a_*$, inclination angle of the disk, photon index $\Gamma$, high-energy cutoff $E_{\rm cut}$, and ionization $\log\xi_{\rm in}$) of the {\tt relxillion\_nk} fit of GRS~1915+105 after the MCMC run. The 2D plots report the 1, 2, and 3$\sigma$ confidence contours.}
    \label{fig:c3b}
\end{figure}

\begin{figure}
	\includegraphics[width=0.95\linewidth,trim=0.0cm 0.0cm 0.0cm 0.0cm,clip]{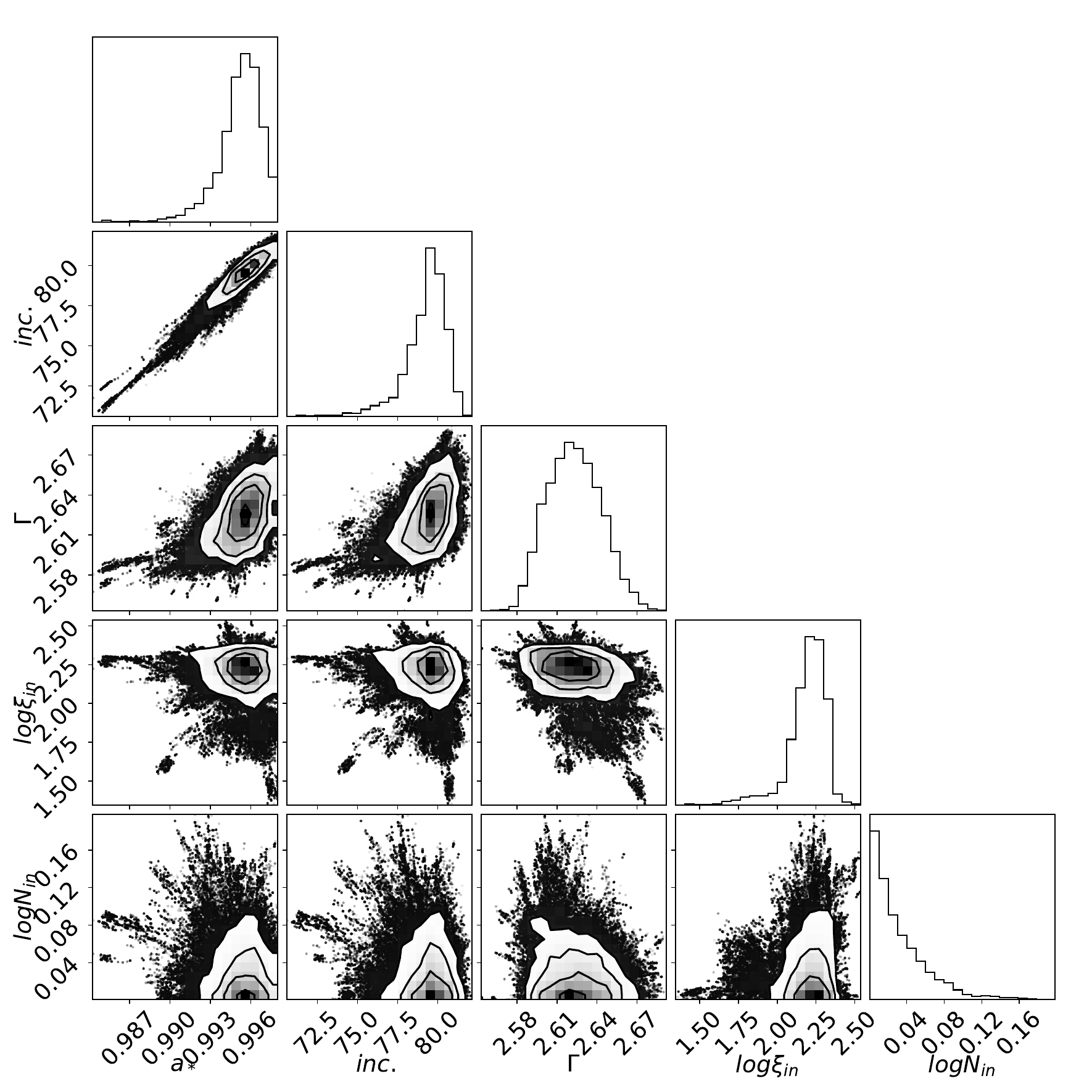}
    \caption{Corner plot for the key-parameter pairs (spin parameter $a_*$, inclination angle of the disk, photon index $\Gamma$, ionization $\log\xi$, and electron density $\log n_{\rm in}$) of the {\tt relxilldgrad\_nk} fit of GRS~1915+105 after the MCMC run. The 2D plots report the 1, 2, and 3$\sigma$ confidence contours.}
    \label{fig:c3c}
\end{figure}

\section{Broadband X-ray spectral analysis}\label{sec:sa}

For the spectral analysis, we use XSPEC v12.9.1 \citep{1996ASPC..101...17A}. We employ the models discussed in Section~\ref{sec:m} with {\tt tbabs}, {\tt powerlaw}, {\tt cutoffpl}, and {\tt xillver}. {\tt tbabs} describes the Galactic absorption~\citep{2000ApJ...542..914W}. It has one parameter, the hydrogen column density $N_{\rm H}$. {\tt powerlaw} describes a power law component and has two parameters: the photon index $\Gamma$ and its normalization. {\tt cutoffpl} describes a power law component with an exponential high-energy cutoff. The model has three parameters: the photon index $\Gamma$, the high-energy cutoff $E_{\rm cut}$, and the normalization of the component. {\tt xillver} describes the reflection spectrum in the rest-frame of the gas~\citep{2013ApJ...768..146G,2014ApJ...782...76G}, so without relativistic effects. It can be used to describe the reflection radiation from some material far from the black hole. The model has six parameters: the photon index $\Gamma$ and the high-energy cutoff $E_{\rm cut}$ of the spectrum of the radiation illuminating the material, the ionization parameter $\xi$ and the iron abundance $A_{\rm Fe}$ of the material, the emission angle $i$, and the normalization of the component.

The differences among {\tt relxill\_nk}, {\tt relxillion\_nk}, and {\tt relxilldgrad\_nk} have already been discussed in Section~\ref{sec:m}. Here we briefly summarize the common features. The models have 13~more parameters in addition to those already discussed in Section~\ref{sec:m}. For the emissivity profile of the reflection spectrum, we employ a broken power law and we have three parameters: the emissivity index of the inner part of the accretion disk $q_{\rm in}$, the breaking radius $R_{\rm br}$, and the emissivity index of the outer part of the accretion disk $q_{\rm out}$. In this work, we assume the Kerr metric, so the spacetime is only characterized by the dimensionless black hole spin parameter $a_*$ (the black hole mass does not directly enter the calculations of the reflection spectrum). The geometry of the accretion disk is described by two parameters: the radial coordinate of the inner edge of the disk $R_{\rm in}$, which we assume at the innermost stable circular orbit and becomes thus a function of $a_*$ when measured in units of gravitational radius $r_{\rm g} = M$, and the radial coordinate of the outer edge of the disk $R_{\rm out}$, which is frozen to its default value $R_{\rm out} = 400~r_{\rm g}$. We have also the inclination angle of the disk with respect to the line of sight of the distant observer, $i$, and the redshift of the source $z$, which is set to zero because we analyze Galactic sources and the Doppler boosting for their relative motion is negligible. The spectrum of the corona illuminating the disk is assumed to be a power law with an exponential high-energy cutoff and we have thus two parameters: the photon index $\Gamma$ and the high-energy cutoff $E_{\rm cut}$. The material of the disk is also characterized by the iron abundance $A_{\rm Fe}$, which is measured in units of iron Solar abundance. Lastly, since the output of these models is the reflection spectrum of the accretion disk and the power law component from the corona, there are two parameters to regulate the normalizations of these two components: the normalization of the model and the reflection fraction $R_{\rm f}$, which regulates the relative intensity between the reflection spectrum and the power law spectrum. If $R_{\rm f}$ is frozen to $-1$, the output of these models is only the reflection component.

First, we fit our data with an absorbed power law. In XSPEC language, the model is {\tt tbabs}$\times${\tt powerlaw}. The plots of the data to the best-fit model ratio are shown in Fig.~\ref{fig:figure1a}, Fig.~\ref{fig:figure1b}, and Fig.~\ref{fig:figure1c}, respectively for GS~1354--645, GRS~1739--278, and GRS~1915+105. We clearly see that the three spectra present strong relativistic reflection features, with a broadened iron line at 5-8~keV and a Compton hump peaked around 20~keV.

We thus add our reflection models to fit the reflection features. When we use {\tt relxill\_nk} and {\tt relxillion\_nk}, the reflection fraction $R_{\rm f}$ is left free in the fit and therefore our XSPEC models are

\vspace{0.2cm}

\noindent {\tt tbabs}$\times${\tt relxill\_nk} ,

\vspace{0.1cm}

\noindent {\tt tbabs}$\times${\tt relxillion\_nk} .

\vspace{0.2cm}

\noindent In the case of {\tt relxilldgrad\_nk}, the high-energy cutoff is frozen to 300~keV, which may be a limitation in the case of low temperature coronae, because our data are up to 50-80~keV. We thus prefer to set the reflection fraction to $-1$ and describe the coronal spectrum with {\tt cutoffpl}. The XSPEC model is

\vspace{0.2cm}

\noindent {\tt tbabs}$\times$({\tt cutoffpl} + {\tt relxilldgrad\_nk}) .

\vspace{0.2cm}

\noindent Even if this choice is not self-consistent with the reflection spectrum calculated by {\tt relxilldgrad\_nk}, it can only provide a better fit than the more conservative choice {\tt tbabs}$\times${\tt relxilliondgrad\_nk} with free reflection fraction.

For the data of GS~1354--645 and GRS~1915+105, we obtain already acceptable fits with a relativistic reflection spectrum. For the \textsl{NuSTAR} data of GRS~1739--278, we still have some residual and we add {\tt xillver} to describe the reflection radiation from some distant cold material. In XSPEC language, the final models for GRS~1739--278 are 

\vspace{0.2cm}

\noindent {\tt tbabs}$\times$({\tt relxill\_nk} + {\tt xillver}) ,

\vspace{0.1cm}

\noindent {\tt tbabs}$\times$({\tt relxillion\_nk} + {\tt xillver}) ,

\vspace{0.1cm}

\noindent {\tt tbabs}$\times$({\tt cutoffpl} + {\tt relxilldgrad\_nk} + {\tt xillver}) .

The results of our fits are summarized in Tab.~\ref{tab:table3}, Tab.~\ref{tab:table4}, and Tab.~\ref{tab:table5}, respectively for GS~1354--645, GRS~1739--278, and GRS~1915+105. $^*$ is used to indicate that the parameter is frozen in the fit. The reported uncertainties are obtained with the {\tt error} command in XSPEC and correspond to the 90\% confidence level for one relevant parameter ($\Delta\chi^2 = 2.71$). If there is no lower/upper uncertainty, it means that the parameter is stuck at one of the boundaries of the allowed range. $q_{\rm in}$ and $q_{\rm out}$ are allowed to vary in the range 0 to 10 in the fit. The maximum value of the black hole spin parameter allowed by the model is 0.998. The iron abundance is allowed to vary in the range 0.5 to 10. Fig.~\ref{fig:figure2}, Fig.~\ref{fig:figure3}, and Fig.~\ref{fig:figure4} show the residuals for the final fits of, respectively, GS~1354--645, GRS~1739--278, and GRS~1915+105. Figs.~\ref{fig:c1a}, \ref{fig:c1b}, and \ref{fig:c1c} show the corner plots for the key-parameters of the fits with, respectively, {\tt relxill\_nk}, {\tt relxillion\_nk}, and {\tt relxilldgrad\_nk} of the GS~1354--645 spectrum after the MCMC run. Figs.~\ref{fig:c2a}, \ref{fig:c2b}, and \ref{fig:c2c} show the same corner plots for GRS~1739--278 and Figs.~\ref{fig:c3a}, \ref{fig:c3b}, and \ref{fig:c3c} are for GRS~1915+105. We postpone the discussion of these results to the next section.


\section{Discussion and conclusions}\label{sec:dc}

The goal of our study is to understand if the available X-ray data of accreting black holes require reflection models with non-trivial ionization and/or electron density profiles and if fitting the data with reflection models with constant ionization and electron density can introduce unacceptably large systematic uncertainties in the estimate of the parameters of the systems. To do this, we have selected three high-quality observations of Galactic black holes with strong relativistic reflection features and we have fit every observation with three models with different assumptions about the ionization and/or electron density profiles. The fits of these observations have already been discussed in \citet{El_Batal_2016} (GS~1354--645), \citet{Miller_2015} (GRS~1739--278), and \citet{Blum_2009} (GRS~1915+105), where these observations were analyzed for the first time, as well as in \citet{Tripathi_2021} and \citet{2020MNRAS.498.3565T}, where these data were re-analyzed with {\tt relxill\_nk}. Here we focus the discussion on the differences among the fits obtained with different assumptions on the ionization and/or electron density profiles.

The first and most important consideration is that, in general, there is no significant difference among the measurements of the parameters of the sources obtained with the three models. In particular, the measurements of the black hole spin parameters $a_*$ and of the inclination angles of the disks $i$ are all consistent.

The estimates of the emissivity profiles are consistent too, and, in particular, we do not find that the inner emissivity index is overestimate by the reflection model with constant ionization and electron density. We note that the emissivity profile of GRS~1739--278 is steep in the inner part of the accretion disk and that the outer emissivity index is $q_{\rm out} \approx 2$-3. This is the typical emissivity profile of a compact corona close to the black hole~\citep{1996MNRAS.282L..53M,2013MNRAS.430.1694D,2020arXiv201207469R}. For the other two sources, GS~1354--645 and GRS~1915+105, we find instead a steep emissivity profile in the inner part of the accretion disk and an almost flat profile for the outer part. This is the emissivity profile that can be expected from an extended corona covering the accretion disk~\citep{2003MNRAS.344L..22M,2012MNRAS.424.1284W,2017MNRAS.472.1932G}. If we model the emissivity profile with a twice broken power law, we find a steep emissivity profile in the inner region ($r <$ a few $r_{\rm g}$), an almost flat profile in the central region (between a few $r_{\rm g}$ and a few hundred $r_{\rm g}$), and $q_{\rm out} \approx 3$ in the outer part~\citep{2022MNRAS.512.2082L}.

\begin{table*}
\caption{AICc values of the fits with {\tt relxill\_nk}, {\tt relxillion\_nk}, and {\tt relxilldgrad\_nk} of the three sources in our study.}
{\renewcommand{\arraystretch}{1.3}
	\begin{tabular}{l c c c}
	    \hline
		\hline
		Source & {\tt relxill\_nk} & {\tt relxillion\_nk} & {\tt relxilldgrad\_nk} \\
		\hline
        GS~1354--645 & 2923.0 & 2911.2 & 2918.8 \\\hline
        	GRS~1739--278 & 1370.7 & 1327.1 & 1367.1 \\\hline
        GRS~1915+105 & 2369.1 & 2366.8 & 2331.0 \\\hline
		\hline
	\end{tabular}
		}\label{tab:tableaicc}
 \end{table*}

From the residuals of the three models (Fig.~\ref{fig:figure2}, Fig.~\ref{fig:figure3}, and Fig.~\ref{fig:figure4}), we do not see any significant difference among the fits of {\tt relxill\_nk}, {\tt relxillion\_nk}, and {\tt relxilldgrad\_nk}. If we compare their $\chi^2$, we see that {\tt relxillion\_nk} provides the best fit for GS~1354--645 and GRS~1739--278, but the value of the ionization index is modest: $\alpha_\xi < 0.4$. In the case of GRS~1915+105, the best fit is provided by {\tt relxilldgrad\_nk}, but just because the data require a higher electron density: indeed $n_{\rm in}$ is stuck at the maximum value allowed by the model, but $\alpha_{\rm n}$ is close to zero and therefore we have an almost constant electron density.

Comparing the minimum of $\chi^2$ of different models is not a particularly robust method to determine which model is favored by the data. A more robust method is the Akaike information criterion (AIC) \citep{1974ITAC...19..716A}. Here we employ the Akaike information criterion corrected for small sample sizes (AICc)~\citep[see, e.g.,][]{bookaicc}, which is more appropriate in our case because the sample size is not large with respect to the number of free parameters. Since we have already the minimum of $\chi^2$ for every model, $\chi^2_{\rm min}$, AICc is straightforward to calculate
\begin{eqnarray}
{\rm AICc} = \chi^2_{\rm min} + 2 N_p + \frac{2 N_p \left( N_p + 1 \right)}{\left( N_b - N_p - 1 \right)} \, ,
\end{eqnarray}
where $N_p$ is the number of free parameters and $N_b$ is the number of bins. Tab.~\ref{tab:tableaicc} shows the values of AICc for every model for the three sources. As a general and empirical rule, we can say that a model with $\Delta{\rm AICc}>5$ (where $\Delta{\rm AICc}$ is the difference between the AIC value of the model and the AIC value of the model with the lowest AIC) is less favored by the data, and that a model with $\Delta{\rm AICc}>10$ can be ruled out and omitted from further consideration~\citep{bookaicc}. With such a criterion, we can confirm that {\tt relxillion\_nk} fits the data of GS~1354--645 and GRS~1739--278 better than {\tt relxill\_nk} and {\tt relxilldgrad\_nk}, and that {\tt relxilldgrad\_nk} is the best model for the data of GRS~1915+105.

The fact that {\tt relxillion\_nk} can fit better the data than {\tt relxill\_nk} is understandable: even if the ionization gradient in the disk is modest in these sources, a non-vanishing ionization gradient is required. The fact that {\tt relxillion\_nk} can provide a lower $\chi^2$ than {\tt relxilldgrad\_nk} for GS~1354--645 and GRS~1739--278 may be related to the limitations imposed by the {\tt xillverD} table, in which $E_{\rm cut}$ is fixed to 300~keV. In particular for the observation of GRS~1739--278, where the fit requires a low value of $E_{\rm cut}$, this may be an important limitation and explain a larger difference between $\chi^2$ of {\tt relxillion\_nk} and {\tt relxilldgrad\_nk} (even if we do not see clear residuals in the bottom panel of Fig.~\ref{fig:figure3} to support such a conclusion). However, for a conclusive answer we should wait for the next version of {\tt relxilldgrad\_nk}, which will use a reflection table with free $E_{\rm cut}$. We note that if we had used a free reflection fraction in {\tt relxilldgrad\_nk} rather than adding {\tt cutoffpl}, as it would have been required in a consistent model, the difference of $\chi^2$ between the models with {\tt relxillion\_nk} and {\tt relxilldgrad\_nk} would have been even larger, because we would have one less free parameter.

We note that the only discrepancy among the reflection models is in the estimate of the high-energy cutoff in the spectrum of the corona. For GS~1354--645 and GRS~1739--278, the value of $E_{\rm cut}$ inferred by {\tt relxill\_nk} is about half the value obtained by {\tt relxillion\_nk}, while the comparison with the best-fit found by {\tt relxilldgrad\_nk} is not straightforward because of the already mentioned point of the {\tt xillverD} table. Such a discrepancy is not present in the fits of GRS~1915+105 presumably because of the very modest ionization gradient required by that observation.

We also note that in the paper describing {\tt relxilldgrad\_nk}~\citep{2021ApJ...923..175A}, we fit a \textsl{NuSTAR} spectrum of the Galactic black hole EXO~1846--031 with our three models ({\tt relxill\_nk}, {\tt relxillion\_nk}, and {\tt relxilldgrad\_nk}), with results very similar to those in the present work. The measurements of the parameters of the sources, and in particular the measurements of the black hole spin parameter, the inclination angle of the disk, and the emissivity profiles, were consistent among the three models, with the only exception of the high-energy cutoff $E_{\rm cut}$ (but for EXO~1846--031 the value found with {\tt relxill\_nk} was higher than the value inferred by {\tt relxillion\_nk}). The lowest $\chi^2$ was obtained in the fit with {\tt relxillion\_nk}, followed by the $\chi^2$ of {\tt relxilldgrad\_nk} and the highest $\chi^2$ was found with {\tt relxill\_nk}. It is remarkable that for EXO~1846--031 the fit with {\tt relxill\_nk} required a Gaussian and was unable to constrain the reflection fraction $R_{\rm f}$, while the fits with {\tt relxillion\_nk} and {\tt relxilldgrad\_nk} did not require any Gaussian and we were able to constrain the reflection fraction or the normalization of {\tt cuttoffpl}.

\citet{2022MNRAS.512..761W} fit the broadband (0.3-50~keV) reflection spectrum of the Seyfert galaxy I~Zwicky~1 and find that an ionization gradient is required, but they do not report the parameter estimates without ionization gradient for a comparison. In their case, the difference between the residuals of the model with and without ionization gradient is clear. Since their analysis includes even the very soft X-ray band, which is not our case here, their fit may be more sensitive to an ionization gradient (see Figs.~\ref{fig:001} and \ref{fig:002}). \citet{2022MNRAS.512..761W} argue that the ionization parameter may fall as $r^{-17/2}$ in the inner part of the accretion disk for a compact corona close to the black hole and as $r^{-3/2}$ at large radii. This is not what we find in our fits with {\tt relxillion\_nk}, where $\alpha_\xi < 0.4$ at 90\% confidence level in all our sources.

To conclude, our study based on a small number of high-quality spectra of Galactic black holes suggests that in many sources the ionization gradient is probably modest and, even if the models with non-vanishing ionization gradient provide better fits, the models with constant ionization provide reliable estimates of the main parameters of the system. However, it is certainly possible that some sources present steep ionization gradients and that their analysis strictly requires models with ionization gradients.


\section*{Acknowledgements}

This work was supported by the National Natural Science Foundation of China (NSFC), Grant No. 11973019, the Natural Science Foundation of Shanghai, Grant No. 22ZR1403400, the Shanghai Municipal Education Commission, Grant No. 2019-01-07-00-07-E00035, and Fudan University, Grant No. JIH1512604. GM acknowledges also the support from the China Scholarship Council (CSC), Grant No. 2020GXZ016647. This work used the data and software provided by the High Energy Astrophysics Science Archive Research Center (HEASARC), which is a service of the Astrophysics Science Division at NASA/GSFC.


\section*{DATA AVAILABILITY}

The \textsl{NuSTAR} and \textsl{Suzaku} raw data analysed in this work are available to download at the HEASARC Data Archive website\footnote{\href{https://heasarc.gsfc.nasa.gov/docs/archive.html}{https://heasarc.gsfc.nasa.gov/docs/archive.html}.}. The reflection models used in this work are available from the corresponding author (C.B.) upon reasonable request and will be soon public at \href{https://github.com/ABHModels}{https://github.com/ABHModels}.



\bibliographystyle{mnras}
\bibliography{Cite} 

\begin{thebibliography}{}
\makeatletter
\relax
\def\mn@urlcharsother{\let\do\@makeother \do\$\do\&\do\#\do\^\do\_\do\%\do\~}
\def\mn@doi{\begingroup\mn@urlcharsother \@ifnextchar [ {\mn@doi@}
  {\mn@doi@[]}}
\def\mn@doi@[#1]#2{\def\@tempa{#1}\ifx\@tempa\@empty \href
  {http://dx.doi.org/#2} {doi:#2}\else \href {http://dx.doi.org/#2} {#1}\fi
  \endgroup}
\def\mn@eprint#1#2{\mn@eprint@#1:#2::\@nil}
\def\mn@eprint@arXiv#1{\href {http://arxiv.org/abs/#1} {{\tt arXiv:#1}}}
\def\mn@eprint@dblp#1{\href {http://dblp.uni-trier.de/rec/bibtex/#1.xml}
  {dblp:#1}}
\def\mn@eprint@#1:#2:#3:#4\@nil{\def\@tempa {#1}\def\@tempb {#2}\def\@tempc
  {#3}\ifx \@tempc \@empty \let \@tempc \@tempb \let \@tempb \@tempa \fi \ifx
  \@tempb \@empty \def\@tempb {arXiv}\fi \@ifundefined
  {mn@eprint@\@tempb}{\@tempb:\@tempc}{\expandafter \expandafter \csname
  mn@eprint@\@tempb\endcsname \expandafter{\@tempc}}}

\bibitem[\protect\citeauthoryear{{Abdikamalov}, {Ayzenberg}, {Bambi}, {Dauser},
  {Garc{\'\i}a}  \& {Nampalliwar}}{{Abdikamalov}
  et~al.}{2019}]{2019ApJ...878...91A}
{Abdikamalov} A.~B.,  {Ayzenberg} D.,  {Bambi} C.,  {Dauser} T.,  {Garc{\'\i}a}
  J.~A.,   {Nampalliwar} S.,  2019, \mn@doi [\apj] {10.3847/1538-4357/ab1f89},
  \href {https://ui.adsabs.harvard.edu/abs/2019ApJ...878...91A} {878, 91}

\bibitem[\protect\citeauthoryear{{Abdikamalov}, {Ayzenberg}, {Bambi}, {Dauser},
  {Garc{\'\i}a}, {Nampalliwar}, {Tripathi}  \& {Zhou}}{{Abdikamalov}
  et~al.}{2020}]{2020ApJ...899...80A}
{Abdikamalov} A.~B.,  {Ayzenberg} D.,  {Bambi} C.,  {Dauser} T.,  {Garc{\'\i}a}
  J.~A.,  {Nampalliwar} S.,  {Tripathi} A.,   {Zhou} M.,  2020, \mn@doi [\apj]
  {10.3847/1538-4357/aba625}, \href
  {https://ui.adsabs.harvard.edu/abs/2020ApJ...899...80A} {899, 80}

\bibitem[\protect\citeauthoryear{{Abdikamalov}, {Ayzenberg}, {Bambi}, {Liu}  \&
  {Zhang}}{{Abdikamalov} et~al.}{2021a}]{2021PhRvD.103j3023A}
{Abdikamalov} A.~B.,  {Ayzenberg} D.,  {Bambi} C.,  {Liu} H.,   {Zhang} Y.,
  2021a, \mn@doi [\prd] {10.1103/PhysRevD.103.103023}, \href
  {https://ui.adsabs.harvard.edu/abs/2021PhRvD.103j3023A} {103, 103023}

\bibitem[\protect\citeauthoryear{{Abdikamalov}, {Ayzenberg}, {Bambi}, {Liu}  \&
  {Tripathi}}{{Abdikamalov} et~al.}{2021b}]{2021ApJ...923..175A}
{Abdikamalov} A.~B.,  {Ayzenberg} D.,  {Bambi} C.,  {Liu} H.,   {Tripathi} A.,
  2021b, \mn@doi [\apj] {10.3847/1538-4357/ac3237}, \href
  {https://ui.adsabs.harvard.edu/abs/2021ApJ...923..175A} {923, 175}

\bibitem[\protect\citeauthoryear{{Akaike}}{{Akaike}}{1974}]{1974ITAC...19..716A}
{Akaike} H.,  1974, IEEE Transactions on Automatic Control, \href
  {https://ui.adsabs.harvard.edu/abs/1974ITAC...19..716A} {19, 716}

\bibitem[\protect\citeauthoryear{{Arnaud}}{{Arnaud}}{1996}]{1996ASPC..101...17A}
{Arnaud} K.~A.,  1996, in {Jacoby} G.~H.,  {Barnes} J.,  eds,  Astronomical
  Society of the Pacific Conference Series Vol. 101, Astronomical Data Analysis
  Software and Systems V. p.~17

\bibitem[\protect\citeauthoryear{{Bambi}}{{Bambi}}{2017a}]{2017bhlt.book.....B}
{Bambi} C.,  2017a, {Black Holes: A Laboratory for Testing Strong Gravity}.
Springer Singapore, \mn@doi{10.1007/978-981-10-4524-0}

\bibitem[\protect\citeauthoryear{{Bambi}}{{Bambi}}{2017b}]{2017RvMP...89b5001B}
{Bambi} C.,  2017b, \mn@doi [Reviews of Modern Physics]
  {10.1103/RevModPhys.89.025001}, \href
  {https://ui.adsabs.harvard.edu/abs/2017RvMP...89b5001B} {89, 025001}

\bibitem[\protect\citeauthoryear{{Bambi}, {C{\'a}rdenas-Avenda{\~n}o},
  {Dauser}, {Garc{\'\i}a}  \& {Nampalliwar}}{{Bambi} et~al.}{2017}]{Bambi_2017}
{Bambi} C.,  {C{\'a}rdenas-Avenda{\~n}o} A.,  {Dauser} T.,  {Garc{\'\i}a}
  J.~A.,   {Nampalliwar} S.,  2017, \mn@doi [\apj] {10.3847/1538-4357/aa74c0},
  \href {https://ui.adsabs.harvard.edu/abs/2017ApJ...842...76B} {842, 76}

\bibitem[\protect\citeauthoryear{{Bambi} et~al.,}{{Bambi}
  et~al.}{2021}]{2021SSRv..217...65B}
{Bambi} C.,  et~al., 2021, \mn@doi [\ssr] {10.1007/s11214-021-00841-8}, \href
  {https://ui.adsabs.harvard.edu/abs/2021SSRv..217...65B} {217, 65}

\bibitem[\protect\citeauthoryear{{Belloni}, {Klein-Wolt}, {M{\'e}ndez}, {van
  der Klis}  \& {van Paradijs}}{{Belloni} et~al.}{2000}]{2000A&A...355..271B}
{Belloni} T.,  {Klein-Wolt} M.,  {M{\'e}ndez} M.,  {van der Klis} M.,   {van
  Paradijs} J.,  2000, \aap, \href
  {https://ui.adsabs.harvard.edu/abs/2000A&A...355..271B} {355, 271}

\bibitem[\protect\citeauthoryear{{Blum}, {Miller}, {Fabian}, {Miller}, {Homan},
  {van der Klis}, {Cackett}  \& {Reis}}{{Blum} et~al.}{2009}]{Blum_2009}
{Blum} J.~L.,  {Miller} J.~M.,  {Fabian} A.~C.,  {Miller} M.~C.,  {Homan} J.,
  {van der Klis} M.,  {Cackett} E.~M.,   {Reis} R.~C.,  2009, \mn@doi [\apj]
  {10.1088/0004-637X/706/1/60}, \href
  {https://ui.adsabs.harvard.edu/abs/2009ApJ...706...60B} {706, 60}

\bibitem[\protect\citeauthoryear{{Brenneman}}{{Brenneman}}{2013}]{2013mams.book.....B}
{Brenneman} L.,  2013, {Measuring the Angular Momentum of Supermassive Black
  Holes}.
Springer New York, \mn@doi{10.1007/978-1-4614-7771-6}

\bibitem[\protect\citeauthoryear{{Burnham} \& {Anderson}}{{Burnham} \&
  {Anderson}}{2002}]{bookaicc}
{Burnham} K.~P.,  {Anderson} D.~R.,  2002, {Model Selection and Multimodel
  Inference}.
Springer New York, \mn@doi{https://doi.org/10.1007/b97636}

\bibitem[\protect\citeauthoryear{{C{\'a}rdenas-Avenda{\~n}o}, {Zhou}  \&
  {Bambi}}{{C{\'a}rdenas-Avenda{\~n}o} et~al.}{2020}]{2020PhRvD.101l3014C}
{C{\'a}rdenas-Avenda{\~n}o} A.,  {Zhou} M.,   {Bambi} C.,  2020, \mn@doi [\prd]
  {10.1103/PhysRevD.101.123014}, \href
  {https://ui.adsabs.harvard.edu/abs/2020PhRvD.101l3014C} {101, 123014}

\bibitem[\protect\citeauthoryear{{Castro-Tirado}, {Brandt}  \&
  {Lund}}{{Castro-Tirado} et~al.}{1992}]{1992IAUC.5590....2C}
{Castro-Tirado} A.~J.,  {Brandt} S.,   {Lund} N.,  1992, \iaucirc, \href
  {https://ui.adsabs.harvard.edu/abs/1992IAUC.5590....2C} {5590, 2}

\bibitem[\protect\citeauthoryear{{Dauser}, {Wilms}, {Reynolds}  \&
  {Brenneman}}{{Dauser} et~al.}{2010}]{2010MNRAS.409.1534D}
{Dauser} T.,  {Wilms} J.,  {Reynolds} C.~S.,   {Brenneman} L.~W.,  2010,
  \mn@doi [\mnras] {10.1111/j.1365-2966.2010.17393.x}, \href
  {https://ui.adsabs.harvard.edu/abs/2010MNRAS.409.1534D} {409, 1534}

\bibitem[\protect\citeauthoryear{{Dauser}, {Garcia}, {Wilms}, {B{\"o}ck},
  {Brenneman}, {Falanga}, {Fukumura}  \& {Reynolds}}{{Dauser}
  et~al.}{2013}]{2013MNRAS.430.1694D}
{Dauser} T.,  {Garcia} J.,  {Wilms} J.,  {B{\"o}ck} M.,  {Brenneman} L.~W.,
  {Falanga} M.,  {Fukumura} K.,   {Reynolds} C.~S.,  2013, \mn@doi [\mnras]
  {10.1093/mnras/sts710}, \href
  {https://ui.adsabs.harvard.edu/abs/2013MNRAS.430.1694D} {430, 1694}

\bibitem[\protect\citeauthoryear{{El-Batal} et~al.,}{{El-Batal}
  et~al.}{2016}]{El_Batal_2016}
{El-Batal} A.~M.,  et~al., 2016, \mn@doi [\apjl] {10.3847/2041-8205/826/1/L12},
  \href {https://ui.adsabs.harvard.edu/abs/2016ApJ...826L..12E} {826, L12}

\bibitem[\protect\citeauthoryear{{Fabian}, {Rees}, {Stella}  \&
  {White}}{{Fabian} et~al.}{1989}]{1989MNRAS.238..729F}
{Fabian} A.~C.,  {Rees} M.~J.,  {Stella} L.,   {White} N.~E.,  1989, \mn@doi
  [\mnras] {10.1093/mnras/238.3.729}, \href
  {https://ui.adsabs.harvard.edu/abs/1989MNRAS.238..729F} {238, 729}

\bibitem[\protect\citeauthoryear{{Fabian}, {Nandra}, {Reynolds}, {Brandt},
  {Otani}, {Tanaka}, {Inoue}  \& {Iwasawa}}{{Fabian}
  et~al.}{1995}]{1995MNRAS.277L..11F}
{Fabian} A.~C.,  {Nandra} K.,  {Reynolds} C.~S.,  {Brandt} W.~N.,  {Otani} C.,
  {Tanaka} Y.,  {Inoue} H.,   {Iwasawa} K.,  1995, \mn@doi [\mnras]
  {10.1093/mnras/277.1.L11}, \href
  {https://ui.adsabs.harvard.edu/abs/1995MNRAS.277L..11F} {277, L11}

\bibitem[\protect\citeauthoryear{{Garc{\'\i}a} \& {Kallman}}{{Garc{\'\i}a} \&
  {Kallman}}{2010}]{2010ApJ...718..695G}
{Garc{\'\i}a} J.,  {Kallman} T.~R.,  2010, \mn@doi [\apj]
  {10.1088/0004-637X/718/2/695}, \href
  {https://ui.adsabs.harvard.edu/abs/2010ApJ...718..695G} {718, 695}

\bibitem[\protect\citeauthoryear{{Garc{\'\i}a}, {Dauser}, {Reynolds},
  {Kallman}, {McClintock}, {Wilms}  \& {Eikmann}}{{Garc{\'\i}a}
  et~al.}{2013}]{2013ApJ...768..146G}
{Garc{\'\i}a} J.,  {Dauser} T.,  {Reynolds} C.~S.,  {Kallman} T.~R.,
  {McClintock} J.~E.,  {Wilms} J.,   {Eikmann} W.,  2013, \mn@doi [\apj]
  {10.1088/0004-637X/768/2/146}, \href
  {https://ui.adsabs.harvard.edu/abs/2013ApJ...768..146G} {768, 146}

\bibitem[\protect\citeauthoryear{{Garc{\'\i}a} et~al.,}{{Garc{\'\i}a}
  et~al.}{2014}]{2014ApJ...782...76G}
{Garc{\'\i}a} J.,  et~al., 2014, \mn@doi [\apj] {10.1088/0004-637X/782/2/76},
  \href {https://ui.adsabs.harvard.edu/abs/2014ApJ...782...76G} {782, 76}

\bibitem[\protect\citeauthoryear{{Gonzalez}, {Wilkins}  \& {Gallo}}{{Gonzalez}
  et~al.}{2017}]{2017MNRAS.472.1932G}
{Gonzalez} A.~G.,  {Wilkins} D.~R.,   {Gallo} L.~C.,  2017, \mn@doi [\mnras]
  {10.1093/mnras/stx2080}, \href
  {https://ui.adsabs.harvard.edu/abs/2017MNRAS.472.1932G} {472, 1932}

\bibitem[\protect\citeauthoryear{{Johannsen} \& {Psaltis}}{{Johannsen} \&
  {Psaltis}}{2013}]{2013ApJ...773...57J}
{Johannsen} T.,  {Psaltis} D.,  2013, \mn@doi [\apj]
  {10.1088/0004-637X/773/1/57}, \href
  {https://ui.adsabs.harvard.edu/abs/2013ApJ...773...57J} {773, 57}

\bibitem[\protect\citeauthoryear{{Kammoun}, {Dom{\v{c}}ek}, {Svoboda},
  {Dov{\v{c}}iak}  \& {Matt}}{{Kammoun} et~al.}{2019}]{2019MNRAS.485..239K}
{Kammoun} E.~S.,  {Dom{\v{c}}ek} V.,  {Svoboda} J.,  {Dov{\v{c}}iak} M.,
  {Matt} G.,  2019, \mn@doi [\mnras] {10.1093/mnras/stz408}, \href
  {https://ui.adsabs.harvard.edu/abs/2019MNRAS.485..239K} {485, 239}

\bibitem[\protect\citeauthoryear{{Kitamoto}, {Tsunemi}, {Pedersen}, {Ilovaisky}
   \& {van der Klis}}{{Kitamoto} et~al.}{1990}]{1990ApJ...361..590K}
{Kitamoto} S.,  {Tsunemi} H.,  {Pedersen} H.,  {Ilovaisky} S.~A.,   {van der
  Klis} M.,  1990, \mn@doi [\apj] {10.1086/169222}, \href
  {https://ui.adsabs.harvard.edu/abs/1990ApJ...361..590K} {361, 590}

\bibitem[\protect\citeauthoryear{{Krimm} et~al.,}{{Krimm}
  et~al.}{2014}]{2014ATel.5986....1K}
{Krimm} H.~A.,  et~al., 2014, The Astronomer's Telegram, \href
  {https://ui.adsabs.harvard.edu/abs/2014ATel.5986....1K} {5986, 1}

\bibitem[\protect\citeauthoryear{{Laor}}{{Laor}}{1991}]{1991ApJ...376...90L}
{Laor} A.,  1991, \mn@doi [\apj] {10.1086/170257}, \href
  {https://ui.adsabs.harvard.edu/abs/1991ApJ...376...90L} {376, 90}

\bibitem[\protect\citeauthoryear{{Liu}, {Liu}, {Bambi}  \& {Ji}}{{Liu}
  et~al.}{2022}]{2022MNRAS.512.2082L}
{Liu} Q.,  {Liu} H.,  {Bambi} C.,   {Ji} L.,  2022, \mn@doi [\mnras]
  {10.1093/mnras/stac616}, \href
  {https://ui.adsabs.harvard.edu/abs/2022MNRAS.512.2082L} {512, 2082}

\bibitem[\protect\citeauthoryear{{Martocchia} \& {Matt}}{{Martocchia} \&
  {Matt}}{1996}]{1996MNRAS.282L..53M}
{Martocchia} A.,  {Matt} G.,  1996, \mn@doi [\mnras] {10.1093/mnras/282.4.L53},
  \href {https://ui.adsabs.harvard.edu/abs/1996MNRAS.282L..53M} {282, L53}

\bibitem[\protect\citeauthoryear{{Martocchia}, {Matt}, {Karas}, {Belloni}  \&
  {Feroci}}{{Martocchia} et~al.}{2002}]{2002A&A...387..215M}
{Martocchia} A.,  {Matt} G.,  {Karas} V.,  {Belloni} T.,   {Feroci} M.,  2002,
  \mn@doi [\aap] {10.1051/0004-6361:20020359}, \href
  {https://ui.adsabs.harvard.edu/abs/2002A&A...387..215M} {387, 215}

\bibitem[\protect\citeauthoryear{{McClintock}, {Shafee}, {Narayan},
  {Remillard}, {Davis}  \& {Li}}{{McClintock} et~al.}{2006}]{McClintock_2006}
{McClintock} J.~E.,  {Shafee} R.,  {Narayan} R.,  {Remillard} R.~A.,  {Davis}
  S.~W.,   {Li} L.-X.,  2006, \mn@doi [\apj] {10.1086/508457}, \href
  {https://ui.adsabs.harvard.edu/abs/2006ApJ...652..518M} {652, 518}

\bibitem[\protect\citeauthoryear{{Miller} et~al.,}{{Miller}
  et~al.}{2013}]{Miller_2013}
{Miller} J.~M.,  et~al., 2013, \mn@doi [\apjl] {10.1088/2041-8205/775/2/L45},
  \href {https://ui.adsabs.harvard.edu/abs/2013ApJ...775L..45M} {775, L45}

\bibitem[\protect\citeauthoryear{{Miller} et~al.,}{{Miller}
  et~al.}{2015a}]{Miller_2015}
{Miller} J.~M.,  et~al., 2015a, \mn@doi [\apjl] {10.1088/2041-8205/799/1/L6},
  \href {https://ui.adsabs.harvard.edu/abs/2015ApJ...799L...6M} {799, L6}

\bibitem[\protect\citeauthoryear{{Miller}, {Reynolds}  \& {Kennea}}{{Miller}
  et~al.}{2015b}]{2015ATel.7612....1M}
{Miller} J.~M.,  {Reynolds} M.~T.,   {Kennea} J.,  2015b, The Astronomer's
  Telegram, \href {https://ui.adsabs.harvard.edu/abs/2015ATel.7612....1M}
  {7612, 1}

\bibitem[\protect\citeauthoryear{{Miniutti}, {Fabian}, {Goyder}  \&
  {Lasenby}}{{Miniutti} et~al.}{2003}]{2003MNRAS.344L..22M}
{Miniutti} G.,  {Fabian} A.~C.,  {Goyder} R.,   {Lasenby} A.~N.,  2003, \mn@doi
  [\mnras] {10.1046/j.1365-8711.2003.06988.x}, \href
  {https://ui.adsabs.harvard.edu/abs/2003MNRAS.344L..22M} {344, L22}

\bibitem[\protect\citeauthoryear{{Mirabel} \& {Rodr{\'\i}guez}}{{Mirabel} \&
  {Rodr{\'\i}guez}}{1999}]{1999ARA&A..37..409M}
{Mirabel} I.~F.,  {Rodr{\'\i}guez} L.~F.,  1999, \mn@doi [\araa]
  {10.1146/annurev.astro.37.1.409}, \href
  {https://ui.adsabs.harvard.edu/abs/1999ARA&A..37..409M} {37, 409}

\bibitem[\protect\citeauthoryear{{Nandra}, {George}, {Mushotzky}, {Turner}  \&
  {Yaqoob}}{{Nandra} et~al.}{1997}]{1997ApJ...477..602N}
{Nandra} K.,  {George} I.~M.,  {Mushotzky} R.~F.,  {Turner} T.~J.,   {Yaqoob}
  T.,  1997, \mn@doi [\apj] {10.1086/303721}, \href
  {https://ui.adsabs.harvard.edu/abs/1997ApJ...477..602N} {477, 602}

\bibitem[\protect\citeauthoryear{{Paul} et~al.,}{{Paul}
  et~al.}{1991}]{1991AdSpR..11h.289P}
{Paul} J.,  et~al., 1991, \mn@doi [Advances in Space Research]
  {10.1016/0273-1177(91)90181-I}, \href
  {https://ui.adsabs.harvard.edu/abs/1991AdSpR..11h.289P} {11, 289}

\bibitem[\protect\citeauthoryear{{Revnivtsev}, {Borozdin}, {Priedhorsky}  \&
  {Vikhlinin}}{{Revnivtsev} et~al.}{2000}]{2000ApJ...530..955R}
{Revnivtsev} M.~G.,  {Borozdin} K.~N.,  {Priedhorsky} W.~C.,   {Vikhlinin} A.,
  2000, \mn@doi [\apj] {10.1086/308386}, \href
  {https://ui.adsabs.harvard.edu/abs/2000ApJ...530..955R} {530, 955}

\bibitem[\protect\citeauthoryear{{Reynolds}}{{Reynolds}}{2014}]{2014SSRv..183..277R}
{Reynolds} C.~S.,  2014, \mn@doi [\ssr] {10.1007/s11214-013-0006-6}, \href
  {https://ui.adsabs.harvard.edu/abs/2014SSRv..183..277R} {183, 277}

\bibitem[\protect\citeauthoryear{{Reynolds} \& {Fabian}}{{Reynolds} \&
  {Fabian}}{2008}]{2008ApJ...675.1048R}
{Reynolds} C.~S.,  {Fabian} A.~C.,  2008, \mn@doi [\apj] {10.1086/527344},
  \href {https://ui.adsabs.harvard.edu/abs/2008ApJ...675.1048R} {675, 1048}

\bibitem[\protect\citeauthoryear{{Reynolds} \& {Nowak}}{{Reynolds} \&
  {Nowak}}{2003}]{2003PhR...377..389R}
{Reynolds} C.~S.,  {Nowak} M.~A.,  2003, \mn@doi [\physrep]
  {10.1016/S0370-1573(02)00584-7}, \href
  {https://ui.adsabs.harvard.edu/abs/2003PhR...377..389R} {377, 389}

\bibitem[\protect\citeauthoryear{{Riaz}, {Ayzenberg}, {Bambi}  \&
  {Nampalliwar}}{{Riaz} et~al.}{2020}]{2020ApJ...895...61R}
{Riaz} S.,  {Ayzenberg} D.,  {Bambi} C.,   {Nampalliwar} S.,  2020, \mn@doi
  [\apj] {10.3847/1538-4357/ab89ab}, \href
  {https://ui.adsabs.harvard.edu/abs/2020ApJ...895...61R} {895, 61}

\bibitem[\protect\citeauthoryear{{Riaz}, {Szanecki}, {Nied{\'z}wiecki},
  {Ayzenberg}  \& {Bambi}}{{Riaz} et~al.}{2021}]{2021ApJ...910...49R}
{Riaz} S.,  {Szanecki} M.,  {Nied{\'z}wiecki} A.,  {Ayzenberg} D.,   {Bambi}
  C.,  2021, \mn@doi [\apj] {10.3847/1538-4357/abe2a3}, \href
  {https://ui.adsabs.harvard.edu/abs/2021ApJ...910...49R} {910, 49}

\bibitem[\protect\citeauthoryear{{Riaz}, {Abdikamalov}, {Ayzenberg}, {Bambi},
  {Wang}  \& {Yu}}{{Riaz} et~al.}{2022}]{2020arXiv201207469R}
{Riaz} S.,  {Abdikamalov} A.~B.,  {Ayzenberg} D.,  {Bambi} C.,  {Wang} H.,
  {Yu} Z.,  2022, \mn@doi [\apj] {10.3847/1538-4357/ac3827}, \href
  {https://ui.adsabs.harvard.edu/abs/2020arXiv201207469R} {925, 51}

\bibitem[\protect\citeauthoryear{{Risaliti} et~al.,}{{Risaliti}
  et~al.}{2013}]{2013Natur.494..449R}
{Risaliti} G.,  et~al., 2013, \mn@doi [\nat] {10.1038/nature11938}, \href
  {https://ui.adsabs.harvard.edu/abs/2013Natur.494..449R} {494, 449}

\bibitem[\protect\citeauthoryear{{Ross} \& {Fabian}}{{Ross} \&
  {Fabian}}{2005}]{2005MNRAS.358..211R}
{Ross} R.~R.,  {Fabian} A.~C.,  2005, \mn@doi [\mnras]
  {10.1111/j.1365-2966.2005.08797.x}, \href
  {https://ui.adsabs.harvard.edu/abs/2005MNRAS.358..211R} {358, 211}

\bibitem[\protect\citeauthoryear{{Shreeram} \& {Ingram}}{{Shreeram} \&
  {Ingram}}{2020}]{2020MNRAS.492..405S}
{Shreeram} S.,  {Ingram} A.,  2020, \mn@doi [\mnras] {10.1093/mnras/stz3455},
  \href {https://ui.adsabs.harvard.edu/abs/2020MNRAS.492..405S} {492, 405}

\bibitem[\protect\citeauthoryear{{Svoboda}, {Dov{\v{c}}iak}, {Goosmann},
  {Jethwa}, {Karas}, {Miniutti}  \& {Guainazzi}}{{Svoboda}
  et~al.}{2012}]{2012A&A...545A.106S}
{Svoboda} J.,  {Dov{\v{c}}iak} M.,  {Goosmann} R.~W.,  {Jethwa} P.,  {Karas}
  V.,  {Miniutti} G.,   {Guainazzi} M.,  2012, \mn@doi [\aap]
  {10.1051/0004-6361/201219701}, \href
  {https://ui.adsabs.harvard.edu/abs/2012A&A...545A.106S} {545, A106}

\bibitem[\protect\citeauthoryear{{Tanaka} et~al.,}{{Tanaka}
  et~al.}{1995}]{1995Natur.375..659T}
{Tanaka} Y.,  et~al., 1995, \mn@doi [\nat] {10.1038/375659a0}, \href
  {https://ui.adsabs.harvard.edu/abs/1995Natur.375..659T} {375, 659}

\bibitem[\protect\citeauthoryear{{Tripathi}, {Zhou}, {Abdikamalov},
  {Ayzenberg}, {Bambi}  \& {Nampalliwar}}{{Tripathi}
  et~al.}{2020a}]{2020PhRvD.102j3009T}
{Tripathi} A.,  {Zhou} B.,  {Abdikamalov} A.~B.,  {Ayzenberg} D.,  {Bambi} C.,
   {Nampalliwar} S.,  2020a, \mn@doi [\prd] {10.1103/PhysRevD.102.103009},
  \href {https://ui.adsabs.harvard.edu/abs/2020PhRvD.102j3009T} {102, 103009}

\bibitem[\protect\citeauthoryear{{Tripathi}, {Liu}  \& {Bambi}}{{Tripathi}
  et~al.}{2020b}]{2020MNRAS.498.3565T}
{Tripathi} A.,  {Liu} H.,   {Bambi} C.,  2020b, \mn@doi [\mnras]
  {10.1093/mnras/staa2618}, \href
  {https://ui.adsabs.harvard.edu/abs/2020MNRAS.498.3565T} {498, 3565}

\bibitem[\protect\citeauthoryear{{Tripathi}, {Zhang}, {Abdikamalov},
  {Ayzenberg}, {Bambi}, {Jiang}, {Liu}  \& {Zhou}}{{Tripathi}
  et~al.}{2021a}]{Tripathi_2021}
{Tripathi} A.,  {Zhang} Y.,  {Abdikamalov} A.~B.,  {Ayzenberg} D.,  {Bambi} C.,
   {Jiang} J.,  {Liu} H.,   {Zhou} M.,  2021a, \mn@doi [\apj]
  {10.3847/1538-4357/abf6cd}, \href
  {https://ui.adsabs.harvard.edu/abs/2021ApJ...913...79T} {913, 79}

\bibitem[\protect\citeauthoryear{{Tripathi}, {Abdikamalov}, {Ayzenberg},
  {Bambi}  \& {Liu}}{{Tripathi} et~al.}{2021b}]{2021ApJ...913..129T}
{Tripathi} A.,  {Abdikamalov} A.~B.,  {Ayzenberg} D.,  {Bambi} C.,   {Liu} H.,
  2021b, \mn@doi [\apj] {10.3847/1538-4357/abf6c5}, \href
  {https://ui.adsabs.harvard.edu/abs/2021ApJ...913..129T} {913, 129}

\bibitem[\protect\citeauthoryear{{Wilkins} \& {Fabian}}{{Wilkins} \&
  {Fabian}}{2012}]{2012MNRAS.424.1284W}
{Wilkins} D.~R.,  {Fabian} A.~C.,  2012, \mn@doi [\mnras]
  {10.1111/j.1365-2966.2012.21308.x}, \href
  {https://ui.adsabs.harvard.edu/abs/2012MNRAS.424.1284W} {424, 1284}

\bibitem[\protect\citeauthoryear{{Wilkins}, {Gallo}, {Costantini}, {Brandt}  \&
  {Blandford}}{{Wilkins} et~al.}{2022}]{2022MNRAS.512..761W}
{Wilkins} D.~R.,  {Gallo} L.~C.,  {Costantini} E.,  {Brandt} W.~N.,
  {Blandford} R.~D.,  2022, \mn@doi [\mnras] {10.1093/mnras/stac416}, \href
  {https://ui.adsabs.harvard.edu/abs/2022MNRAS.512..761W} {512, 761}

\bibitem[\protect\citeauthoryear{{Wilms}, {Allen}  \& {McCray}}{{Wilms}
  et~al.}{2000}]{2000ApJ...542..914W}
{Wilms} J.,  {Allen} A.,   {McCray} R.,  2000, \mn@doi [\apj] {10.1086/317016},
  \href {https://ui.adsabs.harvard.edu/abs/2000ApJ...542..914W} {542, 914}

\bibitem[\protect\citeauthoryear{{Zhang}, {Cui}  \& {Chen}}{{Zhang}
  et~al.}{1997}]{Zhang_1997}
{Zhang} S.~N.,  {Cui} W.,   {Chen} W.,  1997, \mn@doi [\apjl] {10.1086/310705},
  \href {https://ui.adsabs.harvard.edu/abs/1997ApJ...482L.155Z} {482, L155}

\bibitem[\protect\citeauthoryear{{Zhang}, {Abdikamalov}, {Ayzenberg}, {Bambi},
  {Dauser}, {Garc{\'\i}a}  \& {Nampalliwar}}{{Zhang}
  et~al.}{2019a}]{2019ApJ...875...41Z}
{Zhang} Y.,  {Abdikamalov} A.~B.,  {Ayzenberg} D.,  {Bambi} C.,  {Dauser} T.,
  {Garc{\'\i}a} J.~A.,   {Nampalliwar} S.,  2019a, \mn@doi [\apj]
  {10.3847/1538-4357/ab0e79}, \href
  {https://ui.adsabs.harvard.edu/abs/2019ApJ...875...41Z} {875, 41}

\bibitem[\protect\citeauthoryear{{Zhang}, {Abdikamalov}, {Ayzenberg}, {Bambi}
  \& {Nampalliwar}}{{Zhang} et~al.}{2019b}]{2019ApJ...884..147Z}
{Zhang} Y.,  {Abdikamalov} A.~B.,  {Ayzenberg} D.,  {Bambi} C.,   {Nampalliwar}
  S.,  2019b, \mn@doi [\apj] {10.3847/1538-4357/ab4271}, \href
  {https://ui.adsabs.harvard.edu/abs/2019ApJ...884..147Z} {884, 147}

\bibitem[\protect\citeauthoryear{{Zhou}, {Ayzenberg}, {Bambi}  \&
  {Nampalliwar}}{{Zhou} et~al.}{2020}]{2020PhRvD.101d3010Z}
{Zhou} M.,  {Ayzenberg} D.,  {Bambi} C.,   {Nampalliwar} S.,  2020, \mn@doi
  [\prd] {10.1103/PhysRevD.101.043010}, \href
  {https://ui.adsabs.harvard.edu/abs/2020PhRvD.101d3010Z} {101, 043010}

\makeatother
\end{thebibliography}

\bsp	
\label{lastpage}
\end{document}